\documentclass[lettersize,journal]{IEEEtran}
\usepackage{amsmath,amsfonts}
\usepackage{algorithmic}
\usepackage{algorithm}
\usepackage{subfigure} %Para poder hacer subplots
\usepackage{array}
\usepackage{textcomp}
\usepackage{stfloats}
\usepackage{url}
\usepackage{verbatim}
\usepackage{graphicx}
\usepackage{cite}
\usepackage{gensymb}
\usepackage{xcolor,soul,framed} %,caption
\usepackage{dcolumn}% Align table columns on decimal point
\usepackage{bm}% bold math
\usepackage{subfigure} %Para poder hacer subplots
\usepackage{pict2e}
\usepackage{epstopdf}
\usepackage{arydshln}
\usepackage{lipsum}
\usepackage{booktabs}
\usepackage{float}
\usepackage[hidelinks]{hyperref}

\begin{document}

\title{Enabling FR2-5G Communication with Dielectric OAM Transmitarrays}

\author{Miguel Á. Balmaseda-Márquez, Juan E. Galeote-Cazorla, Álvaro Liébana-Bolívar, Alejandro Ramírez-Arroyo, Carlos Molero Jiménez \IEEEmembership{Member, IEEE}, J.F. Valenzuela-Valdés.

\thanks{
This work has been supported by grant PID2024-157242OB-C44 and PID2020-112545RB-C54 funded by MCIN/AEI/10.13039/501100011033 and by the European Union NextGenerationEU/PRTR. It has also been supported by grants PDC2023-145862-I00, FPU22/03392 and FPU21/02219 funding by MCIN/AEI/10.13039/501100011033 and by European Union NextGenerationEU/PRTR. It has also been funded by, in part, Grant PID2024-155167OA-I00 funded by MICIU/AEI/10.13039/501100011033/FEDER, UE,  and in part by Consejería de Universidad, Investigación e Innovación of Junta de Andalucía through grant EMERGIA 23-00235.}

\thanks{Miguel Á. Balmaseda, Juan E. Galeote-Cazorla, Álvaro Liébana-Bolíbar and J. F. Valenzuela-Valdés are with Dept. of Signal Theory, Telematics and Communications and with the Centre for Information and Communication Technologies (CITIC-UGR), University of Granada, Granada, Spain (e-mail: migbalmar@ugr.es, juane@ugr.es, aliebana2003@correo.ugr.es, juanvalenzuela@ugr.es)}
\thanks{Carlos Molero Jimenez is with Dept. of Electronic and Electromagnetism, Faculty of Physics, University of Seville, 41012, Seville, Spain (e-mail: cmolero1@us.es)}
\thanks{Alejandro Ramírez-Arroyo is with the Department of Electronic Systems, Aalborg University (AAU), 9220 Aalborg, Denmark (e-mail: araar@es.aau.dk)}
}

% The paper headers
\markboth{Balmaseda-Márquez \MakeLowercase{\textit{et al.}}: Enabling FR2-5G Communication with Dielectric OAM
Transmitarrays}%
{Balmaseda-Márquez \MakeLowercase{\textit{et al.}}: Enabling FR2-5G Communication with Dielectric OAM
Transmitarrays}

\maketitle

\begin{abstract}

This paper investigates the potential of near-field (NF) indoor communications in the FR2 frequency bands using fully dielectric structures to generate orbital angular momentum (OAM) waves. All-dielectric platforms based on distributions of T-shaped unit cells are employed for this purpose. The unit cell design is based on a circuital approach and analytical formulations, where phase shifts necessary for OAM generation are achieved by varying the dielectric-to-air ratio within the structure. Based on this unit cell, a set of transmitarrays (TAs) are designed to produce specific OAM modes. These TAs are fabricated in-house using stereolithographic 3D printing and experimentally tested. The tests evaluate two key features of OAM beams: the orthogonality of distinct vortex modes, as characterized by their electric field distributions, and their object-avoidance capability, enabled by the central null characteristic of the wavefront. In addition, a field-test within an indoor environment is conducted emulating a real wireless system. A bit error rate lower than 10\textsuperscript{$-$6} is observed for solidary modes in Tx and Rx, whereas orthogonal modes produce an increment in 4 order of magnitude. The obtained results reveal that the prototype is suitable for short-range scenarios, enabling techniques such as OAM-multiplexation or physical-layer security thanks to the effective orthogonality between modes.
\end{abstract}


\section{Introduction}

In recent years, significant progress has been made in the development and application of devices capable of generating vortex beams, driven by their potential for advanced communication systems and encryption technologies \cite{Shen2022}. Vortex beams, which carry orbital angular momentum (OAM), were first proposed by Allen \textit{et al.} in 1992 \cite{Allen1992}. Among their unique characteristics, two stand out in comparison to conventional waveforms: 
on the one hand, the existence of an \emph{infinite} number of orthogonal topological charges or vorticity orders \cite{Torres2011}; on the other hand their self-reconstruction or self-healing \cite{Bouchal1998}. These properties make vortex beams particularly attractive for modern communication systems. They enable novel approaches to information encryption using reconfigurable hardware platforms \cite{Fang2020}, and support robust signal propagation in complex environments, including scenarios involving physical obstructions \cite{Cheng2024}.

\begin{figure}[t]
    \centering
    {\includegraphics[width=1\columnwidth]{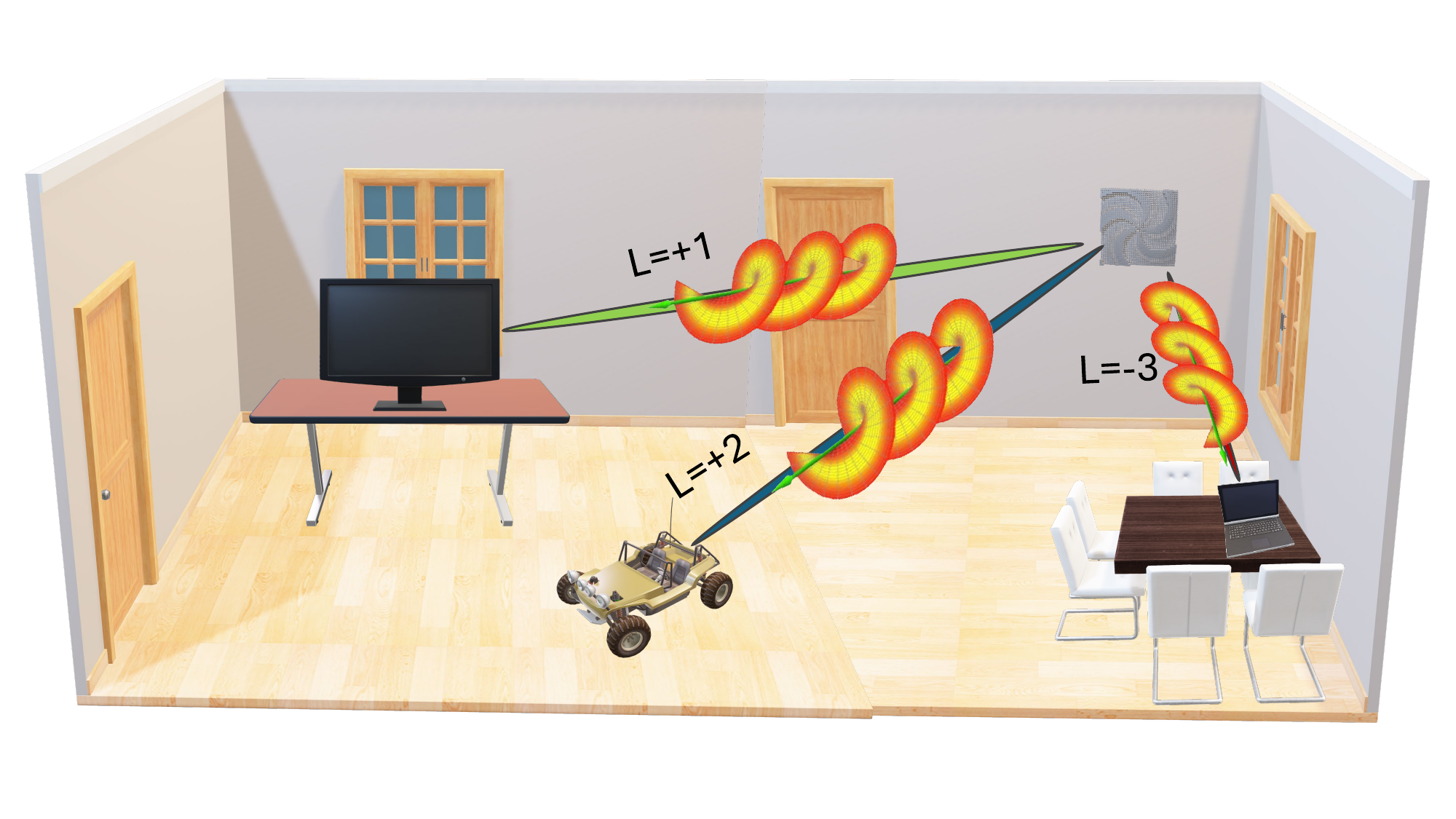}}
    \caption{Indoor scenario with a emitting TA with a multimodal frequency range for OAM communications. Through only one emitter it is possible to transmit several modes to a high number of objectives.}
    \label{fig:OAM_indoors}
 \end{figure}

The generation of OAM-based vortex beams can be achieved using either multiple-antenna configurations or single-platform implementations. Among the most widely utilized multi-antenna systems is the \emph{Uniform Circular Array} (UCA) \cite{Mohammadi2010}, which can be configured with a single circular ring of elements \cite{Edfors2012} or multiple concentric rings to enhance performance \cite{Opare2014}. A primary limitation of UCA-based systems lies in their sensitivity to misalignment between the transmitting (TX) and receiving (RX) terminals \cite{Lyu2023}. To address these alignment challenges and to improve beam directivity, hybrid configurations that combine UCAs with parabolic reflectors have been proposed and experimentally validated \cite{Nguyen2016, Wu2024}.

However, these systems tend to be bulky and heavy, which limits their suitability for mobile applications and increases system complexity due to the need for an external power supply. To mitigate these limitations, low-profile devices based on single-platform architectures and externally fed by incident waves have been developed. Notable examples include spiral phase plates \cite{Massari2015, Zhang2017}, which consist of a slab with a spiral-shaped surface. These structures can operate in both transmission and reflection modes \cite{Papathanasopoulos2021}, but their performance is highly sensitive to fabrication inaccuracies, with imperfections potentially degrading their operational effectiveness \cite{Campbell2012}.

A more versatile approach to OAM beam generation involves the use of metasurface platforms \cite{Quevedo2019}. A conventional metasurface consists of a periodic arrangement of subwavelength cells, each of which can be independently controlled to manipulate the phase, amplitude, or polarization of an incident wave \cite{Epstein2016}. Similar to UCAs, metasurfaces can generate vortex beams; however, the additional degrees of freedom provided by metasurfaces allow for precise tuning of the OAM mode at a single frequency. This enables the generation of multiple OAM states using more compact and cost-effective implementations \cite{Ren2019, Xu2016, Li2021, He2021, Yuan2023}. As an advancement of this concept, fully dielectric metasurfaces have emerged as a low-cost and straightforward solution for OAM beam generation, offering simplified fabrication and improved integration potential \cite{Wang2023_OAM, Odit2019, Lin2020, Ullah2025}.

Nevertheless, conventional metasurfaces are typically limited to generating a single OAM mode at a time, which can be restrictive for applications requiring dynamic mode switching or simultaneous multi-mode operation. To overcome this limitation, programmable metastructures have been introduced as a next-generation solution \cite{Ma2017}. These fully reconfigurable platforms, when combined with the inherent orthogonality of OAM modes, enable multiplexed operation across multiple channels—offering significant advantages for high-capacity communication systems.

Due to the increasing demands imposed by emerging communication systems, advancements have not been limited to hardware enhancements and OAM approaches; concurrently, new frequency bands have also been explored and allocated to accommodate the growing performance requirements. In this context, novel spectrum allocations have been complemented by the development of advanced enabling technologies—most notably those designed to support higher bandwidth and increased link density per antenna, as exemplified by 5G networks \cite{Xiao2017, Rappaport2017, Agiwal2016, Juane2025}. These requirements are particularly pressing in indoor environments characterized by high device density \cite{Gupta2015, Chandra2015, Li2018, Venugopal2016, Al-Samman2018, Alex2024}. In such scenarios, the transmission efficiency between TX and RX terminals has been evaluated using Bessel beams, which are known for their non-diffracting behavior within a limited propagation region \cite{Heebl2016, Kollarova2007, Vetter2019}. Similar non-diverging characteristics can be achieved in OAM-carrying waves through appropriate beam-shaping techniques \cite{Xu2021, Torcolacci2023}. Moreover, OAM beams offer an additional layer of security for communication systems by enabling information encryption through the modulation of vortex orders \cite{Fang2020}.

This work presents a preliminary study on the near-field (NF) indoor communication capabilities of orbital angular momentum (OAM) waves (see Fig.~\ref{fig:OAM_indoors}) \cite{Wang2023-NF, Lee2021-NF, Zhang2023-NF, Liang2025-NF}. To this end, a series of fully dielectric metasurfaces have been designed, fabricated, and experimentally tested in a controlled laboratory environment. These metasurfaces emulate a reconfigurable system, with each device engineered to generate a specific OAM mode, thereby enabling experimental verification of the orthogonality between vortex orders. The metasurface design is based on a circuit model derived from the homogenization of the unit cell, following a methodology similar to that proposed in \cite{Balmaseda2024}. In the present configuration, a T-shaped dielectric block is embedded within an air-filled square unit cell, which enables the realization of single-layer prototypes—offering a simplified alternative to the multi-layer approach described in \cite{Balmaseda2024}. A comprehensive analysis of the circuit model's limitations is conducted, and the model is subsequently used to guide the metasurface design process.

Experimental evaluations focus on two key aspects. First, the object-avoidance capability of OAM waves is investigated, taking advantage of the central null in their wavefront structure \cite{Jofre2024-OAM, Alison2011-OAM}. Second, mode multiplexing at a single operating frequency is demonstrated. Additionally, a field-test campaign is conducted to assess the suitability of the prototype within a NF indoor system evaluating performance parameters such as error vector magnitude (EVM), modulation error rate (MER) and bit error rate (BER). This ability to transmit multiple data streams simultaneously enhances the potential of OAM-based systems for secure, encrypted indoor communication, positioning them as promising candidates for integration into current 5G and emerging 6G frequency bands \cite{Wang2023, Yadanar2025, Ali2021, Ali2023}.

This work is organized as follows. Section II summarizes the dielectric unit‑cell designs together with their characterization. An analytical expression is introduced to evaluate the effective permittivity, and a corresponding circuit model is employed to emulate the unit‑cell behavior. In Section III, the limitations of this model are examined, with particular attention to its validity over frequency and the impact of oblique incidence. Then, Section IV and V is left to conduct several experiments, concerning the self-healing and self reconstruction via OAMs, and conducted field test of OAM-based communications. Finally, the conclusions are presented. 

\begin{figure}[t!]
\centering
\subfigure[]{\includegraphics[width=0.47\columnwidth]{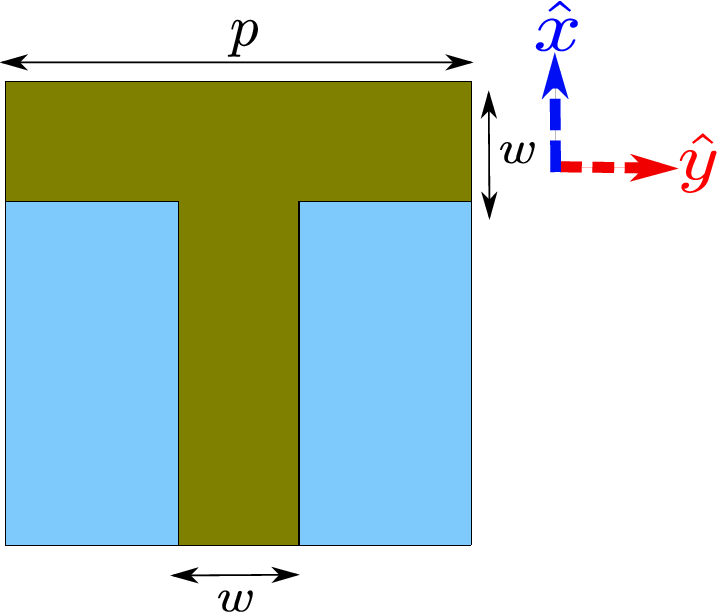}\label{T_front}}
\subfigure[]{\includegraphics[width=0.49\columnwidth]{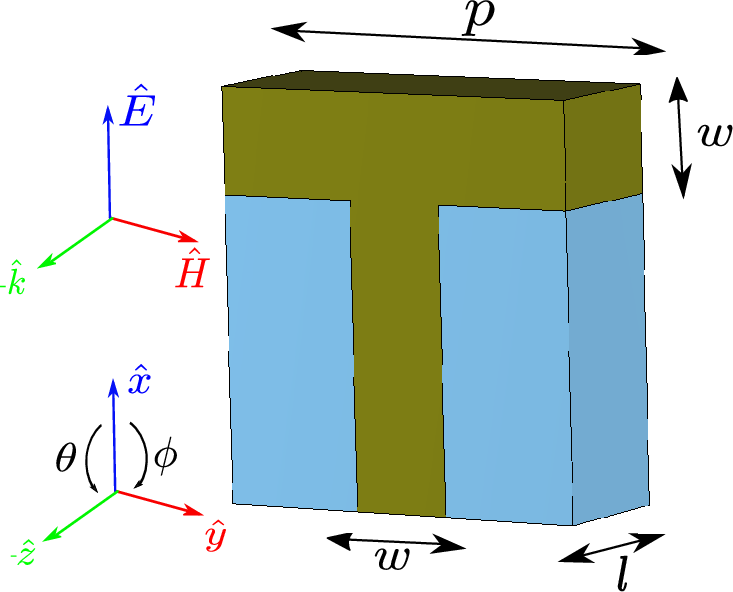}\label{T_perfil}}
\subfigure[]{\includegraphics[width=0.5\columnwidth]{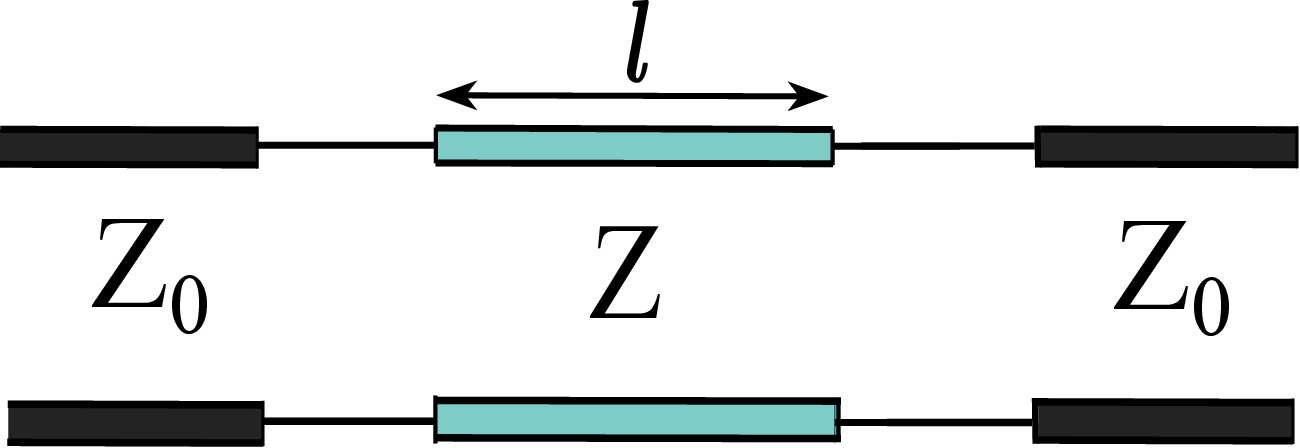}\label{Circuito_simple}}
\caption{(a) Front view of the T-shaped unit cell (b)  Perspective view of the unit cell. Blue color corresponds to air and the brown color to the dielectric material (c) Circuital equivalent model for a T-shaped unit cell of length $l$.}
 \end{figure}

\section{Characterization of the unit cell. Circuit model}

Before the definition of the metasurface, we proceed with the description of its corresponding basic unit, or unit cell. As was aforementioned and as can be visualized in Figs.~\ref{T_front} - \ref{T_perfil}, it consists of a square unit cell with side length (periodicity) $p$ and width $l$. The cell includes a T-shaped dielectric block with permittivity $\varepsilon_{\text{r}}$. The remaining space of the cell is filled by air. The proportion of dielectric with respect the cell size is estimated through the \emph{filling factor} ($\chi$), defined as the ratio between the dielectric surface and the total surface of the transverse section of the unit cell:

\begin{equation}
\label{eq:chi}
\chi = \frac{w(2p-w)}{p^2}\,.
\end{equation}

\begin{figure}[t!]
    \centering
    \subfigure[]{\includegraphics[width=1\columnwidth]{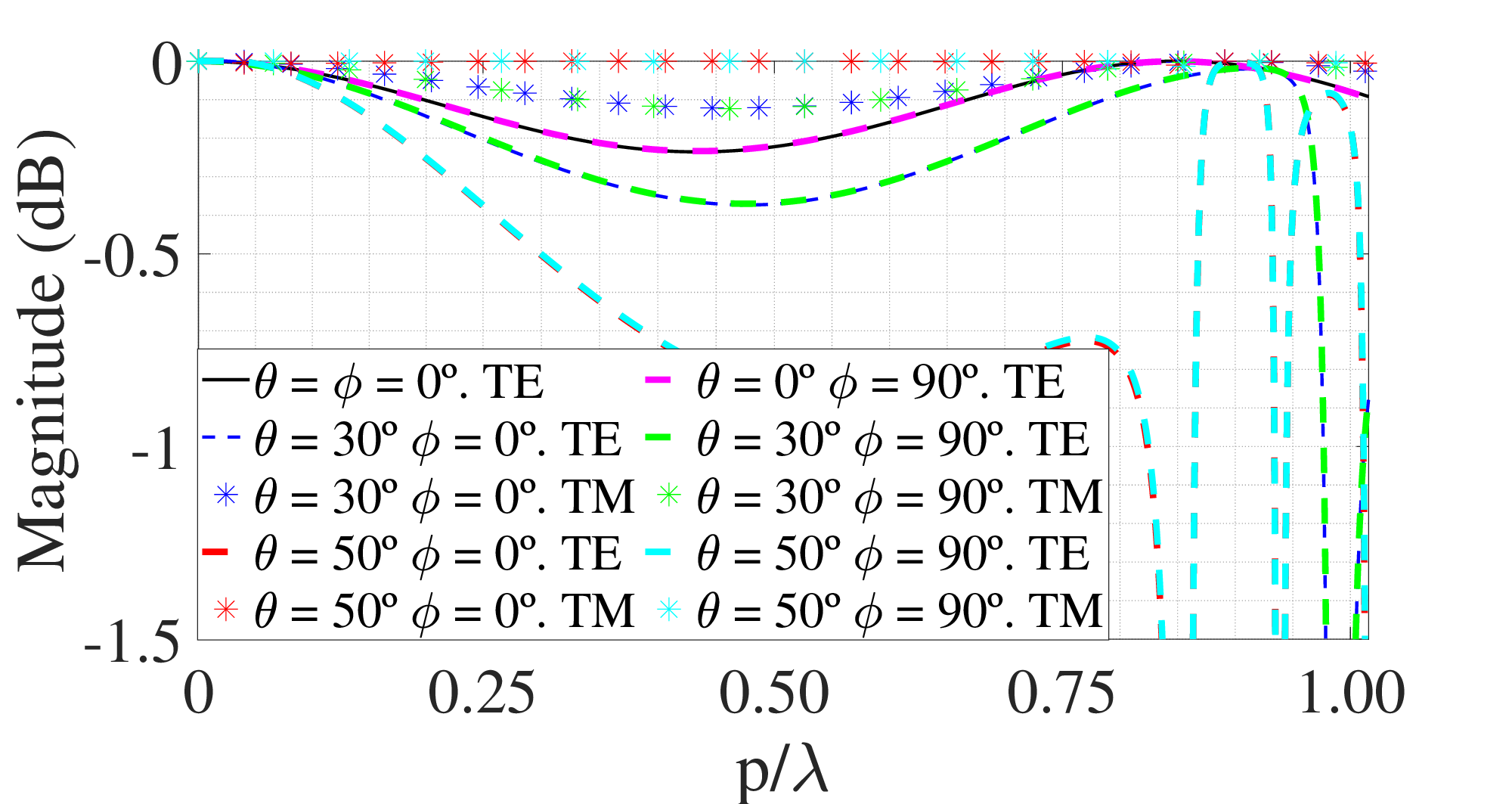}\label{Obl_dB_2.6}}
    \subfigure[]{\includegraphics[width=1\columnwidth]{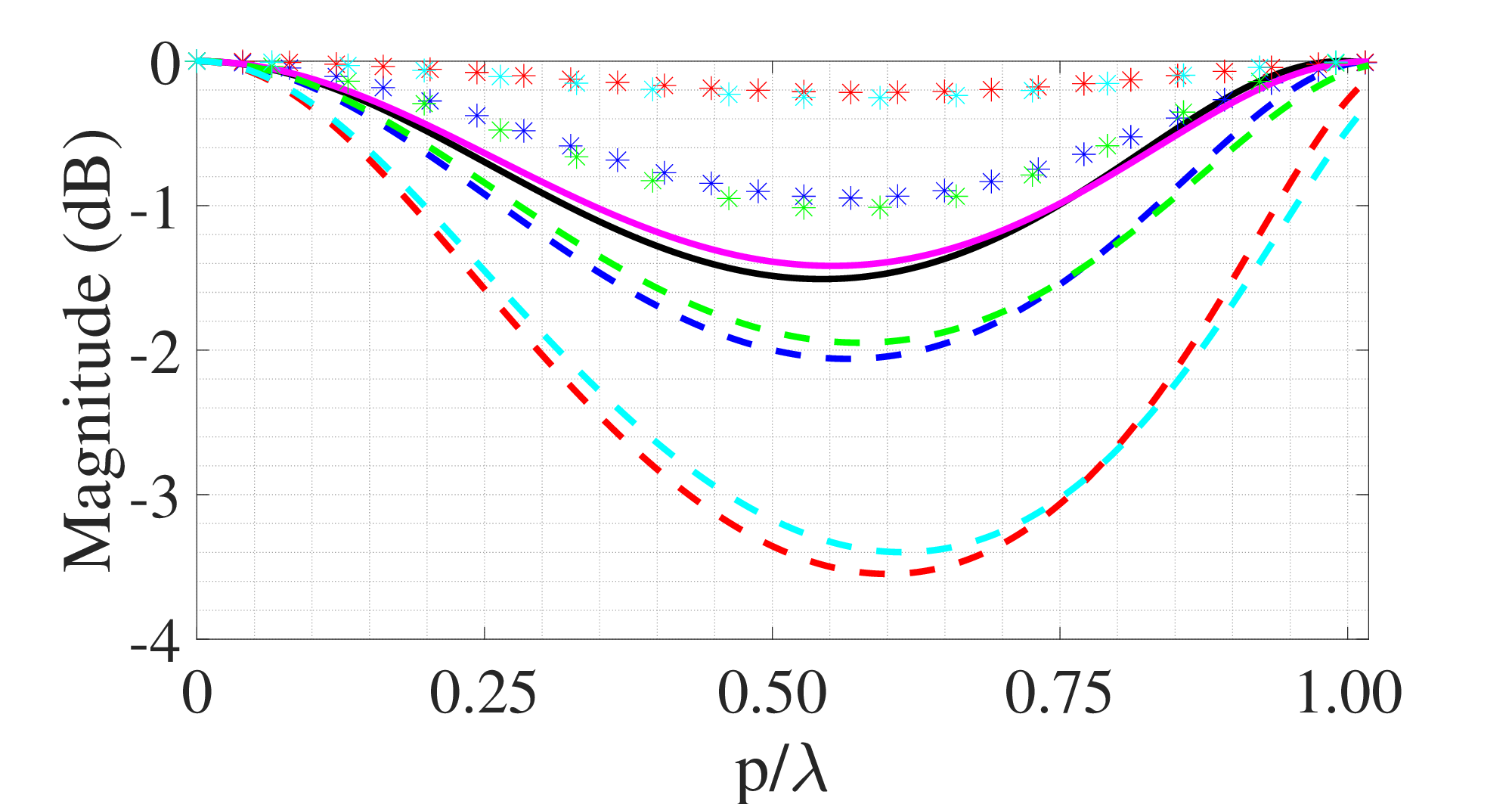}\label{Obl_dB_8}}
    \caption{Simulations to study the oblique incidence (TE and TM incidence) for a T-shaped cell with a $\chi = 0.45$. (a) Transmission coefficient for $\varepsilon_{\text{r}}=2.6$  
    (b) Transmission coefficient for $\varepsilon_{\text{r}}=8$.}
\end{figure}

The filling factor takes values within the interval $\chi = [0, 1]$. The case of $\chi \approx 0$ indicates a large amount of air in the cell, thus becoming an effective medium with permittivity $\varepsilon_{\text{r, eff}} \approx 1$. Similarly, when $\chi \approx 1$, the dielectric-to-air ratio becomes large, causing the structure to behave as an effective medium with $\varepsilon_{\text{r, eff}} \approx \varepsilon_{\text{r}}$. Intermediate $\chi$ values are related to effective media, or \emph{effective permittivity} values  $\varepsilon_{\text{r, eff}}$ between the interval $\varepsilon_{\text{r, eff}} \in [1, \varepsilon_{\text{r}}]$. An extensive study about the evaluation of $\varepsilon_{\text{r, eff}}$ has been done in \cite{Balma_EuCAP-2025}. After a parametric sweep and with a semi-analytical method the effective permittivity behaves as a function of $\chi$ as 
\begin{equation}\label{eq:T_shape}
\varepsilon_{\text{r,eff}} (\varepsilon_{\text{r}}, \chi) =\left(\varepsilon_{\text {r}}-1\right) \cdot \chi^{b}+1 
\end{equation}
where $b$ represents the growth rate of the function and is equal to $b = 1.289$.

As previously discussed, the effective permittivity is associated with an equivalent homogeneous medium that emulates the electromagnetic response of the original unit cell. This homogenized medium can be treated as a uniform dielectric block, and the wave propagation within it can be accurately described in terms of a transmission line model such as that illustrated in Fig.~\ref{Circuito_simple}. In this representation, the incident, reflected, and transmitted plane waves are modeled by input and output transmission lines with a characteristic impedance of $Z_0~=~120\pi\,\Omega$ (assuming normal incidence), corresponding to free space. Otherwise the line with characteristic impedance $Z$ represents the propagation inside the effective medium. Its length is set equal to the width of the unit cell, denoted by $l$. The corresponding propagation constant and characteristic impedance associated with the line are estimated via the effective permittivity as

\begin{equation}\label{beta_eq}
\beta = \sqrt{\varepsilon_{\text{r, eff}}} \frac{2 \pi f}{c} 
\end{equation}
\begin{equation}\label{Z_eq}
Z = \dfrac{Z_{0}}{\sqrt{\varepsilon_{\text{r,eff}}}} 
\end{equation}
with $f$ and $c$ being the operation frequency and the speed of light respectively.

\begin{figure}[t!]
    \centering
    \subfigure[]{\includegraphics[width=0.65\columnwidth]{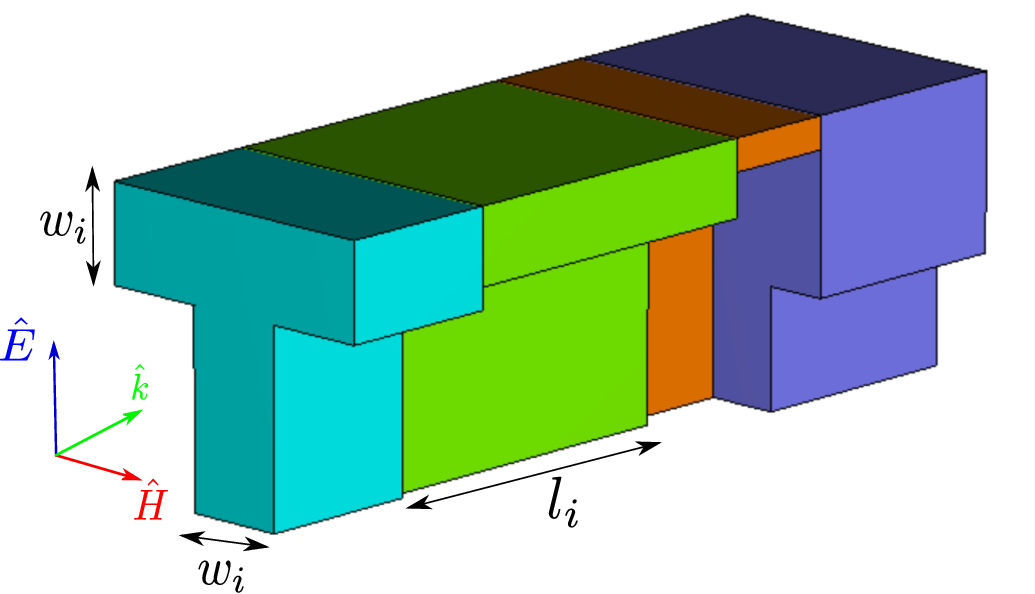}\label{Simul_1}}
    \subfigure[]{\includegraphics[width=0.8\columnwidth]{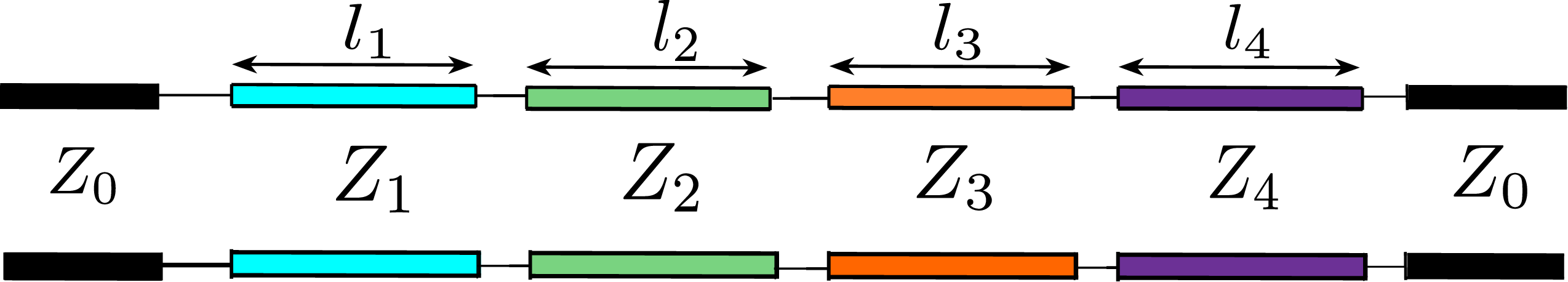}\label{Circ_cuad}}
    \caption{(a) Unit cell formed by 4 different concatenated T-blocks. Periodicity $p = 2.7\,$mm. (b) Equivalent circuit for the concatenated T-blocks whose parameters be seen in Table \ref{table_values}.}
\end{figure}

\subsection{Circuit-model validation. Single and stacked T-blocks}

This section is devoted to conduct a comprehensive analysis of the robustness of the model, both under normal and oblique incidence. TE and TM polarizations are considered. The $E$-planes are fixed at $\phi = 0$\textdegree \ or $\phi = 90$\textdegree, while $\theta$ is varied discretely to evaluate the model performance across different incidence angles. The expression of the input/output characteristic impedance for the equivalent circuit can be expressed as
\begin{align}\label{Oblique_TE}
Z_{0} &= \frac{120\pi}{\cos(\theta)} \hspace{12mm} \text{TE incidence} \\
\label{ObliqueTM} Z_{0} &= 120\pi \cos(\theta) \hspace{5mm} \text{TM incidence}
\end{align}
by assuming free space as input/output media.

A first analysis is focused on the cell in Figs.~\ref{T_front}-\ref{T_perfil}, formed by a single T-block. Two situations are regarded: one involving a T-block with $\varepsilon_{\text{r}} = 2.6$; a second one  with $\varepsilon_{\text{r}} = 8$. A frequency sweep in terms of $p/\lambda$ has been done to check how the structure behaves when $\theta$ goes from $0\degree$ (normal incidence) up to $50\degree$.   

Fig.~\ref{Obl_dB_2.6} shows the results for the transmission coefficient for TE and TM incidence for $\varepsilon_{\text{r}} = 2.6$. The periodicity of the cell is set at $p = 2.7\,$mm and the filling factor $\chi = 0.45$. It can be seen that small differences appear in the results from angles $\theta = 0\degree-30\degree$. TM incidence behaves more similar to normal incidence, even at $\theta = 50$\textdegree. TE incidence otherwise deteriorates beyond $\theta = 30$\textdegree. Notice that the degradation is substantial for $\theta = 50$\textdegree\, beyond $p/\lambda \ge 0.75$. Fig.~\ref{Obl_dB_8} shows the results for $\varepsilon_{\text{r}} = 8$. The behavior is still acceptable when the incidence is of TM nature, and again deteriorates for $\theta = 50$\textdegree \ in the TE case. Now, the degradation is manifested by a substantial reduction of the transmission coefficient, reaching values below $-3\,$dB at some frequencies. Thus, the close inspection of both cases leads us to define a prudential angular limit in $\theta = 30$\textdegree.\ This will be the threshold where the cell is robust with the incidence angle, and where the circuit model can be employed for the design.  

\begin{figure}[]
    \centering
    \subfigure[]{\includegraphics[width=1\columnwidth]{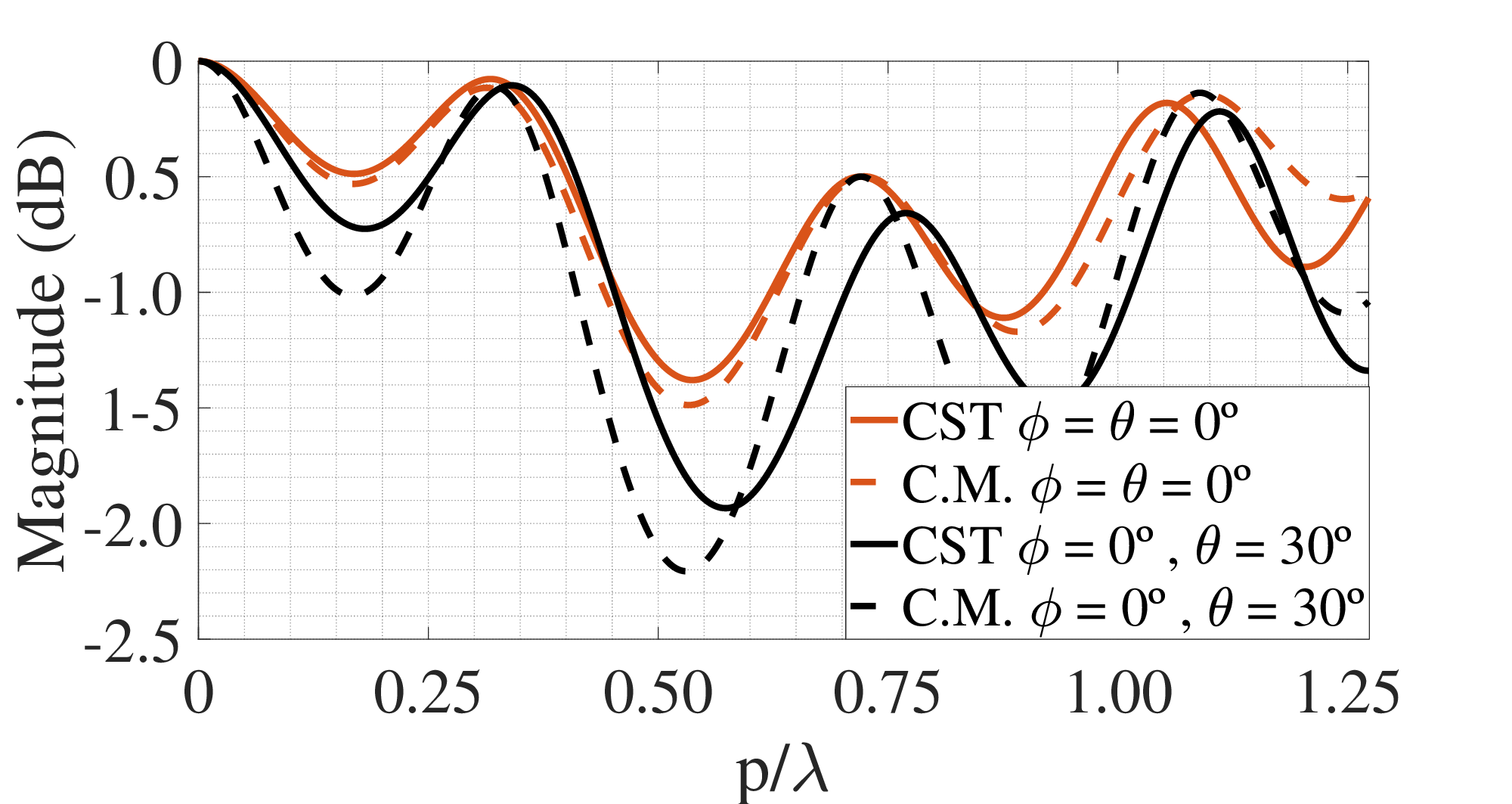}\label{S12_dB}}
    \subfigure[]{\includegraphics[width=1\columnwidth]{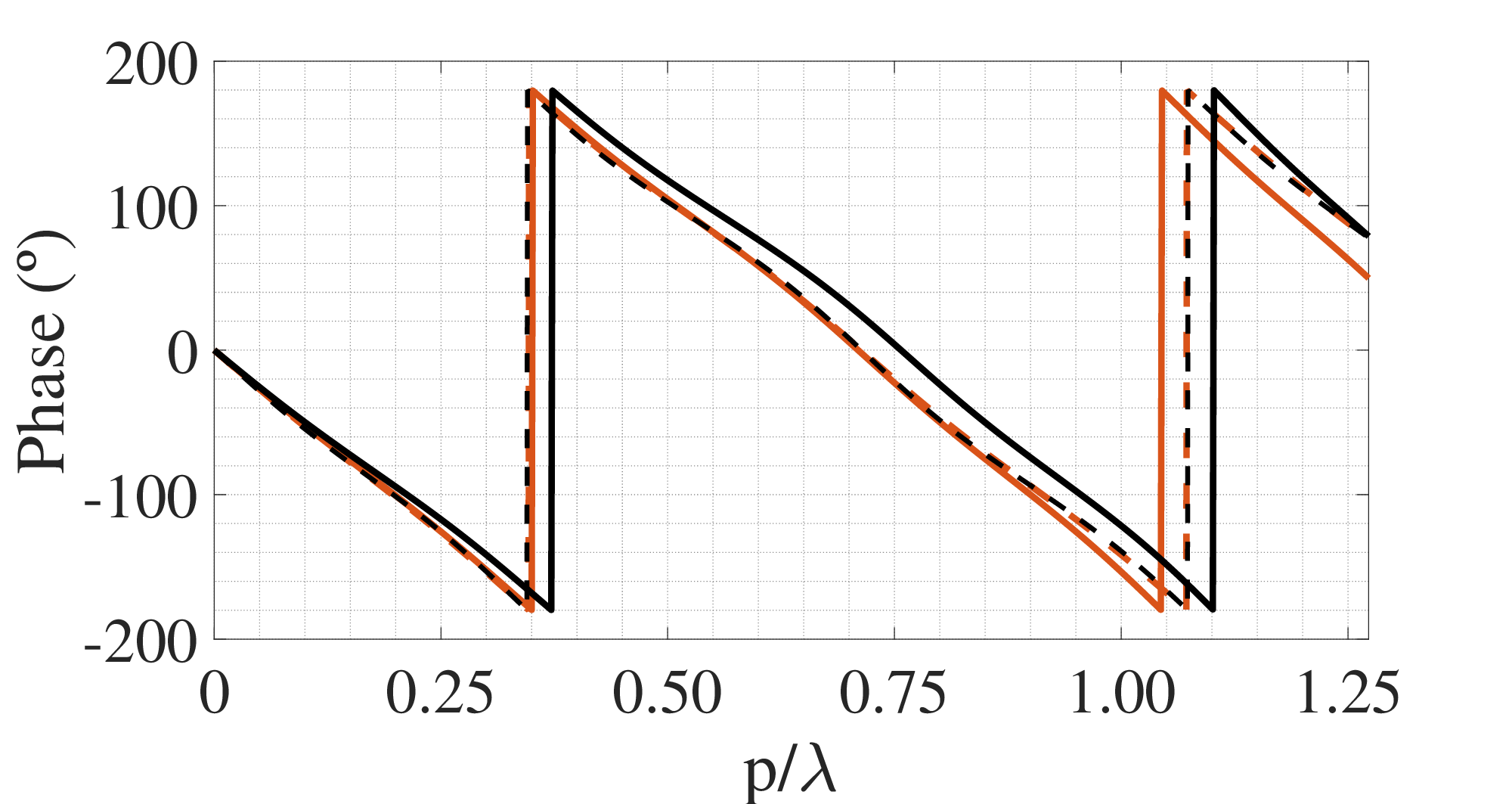}\label{S12_Phase}}
    \caption{Scattering parameters for the cascade of several cells (a) Magnitude of the transmission coefficient (b) Phase of the transmission coefficient.}
 \end{figure}

As a second test and in order to confirm the validity of the model up to $\theta = 30$\textdegree, we conducted an experiment involving a unit cell formed by series of concatenated T-shaped blocks, as sketched in Fig.~\ref{Simul_1}. Each of the T-blocks has different filling factors $\chi_{i}$ and permittivity values $\varepsilon_{\text{r}, i}$. The specific structure parameters of the whole cell are presented in Table~\ref{table_values}. 
\begin{table}[t]
\centering
\caption{Structure parameters of the unit cell in Fig.~\ref{Simul_1}}
\label{table_values}
\begin{tabular}{|c|c|c|c|c|c|}
\hline
\textbf{Block $i$} & $l_{i}$ (mm) & $w_{i}$ (mm)  &\textbf{$\chi_{i}$} & \textbf{$\varepsilon_{\text{r}, i}$} & $Z_{i}\,(\Omega)$ \\ \hline
1 & 1.5 & 0.9 & 0.56 & 6.0 & 206.2 \\ \hline
2 & 3.0 & 0.7 & 0.45 & 3.0 & 287.7 \\ \hline
3 & 1.0 & 0.3 & 0.21 & 8.0 & 271.0 \\ \hline
4 & 2.0 & 1.6 & 0.83 & 2.0 & 281.7 \\ \hline
\end{tabular}
\end{table}
Each T-block can be represented as an individual transmission-line section. The whole cell therefore admits to be represented by four transmission lines connected in parallel, as sketched in Fig.~\ref{Circ_cuad}. This transmission-line model emulates the whole dielectric composite. The characteristic impedance $Z_{i}$ and propagation constants related to each of the lines are estimated by using \eqref{beta_eq}-\eqref{Z_eq}. 
The impedance in free space, $Z_{0}$, for the input/output lines are estimated again according to \eqref{Oblique_TE}-\eqref{ObliqueTM}.

%Moreover, the calculation of $Z_{i}$ in $\eqref{Z_eq}$ takes that into account. 
Figs.~\ref{S12_dB}-\ref{S12_Phase} represent the magnitude and phase of the transmission coefficient of the structure respectively. Results computed by CST and by our circuit model (C.M.) are represented in the same plots. In addition, the plots include results for normal incidence ($\theta = 0$\textdegree, $\phi = 0$\textdegree) and oblique incidence ($\theta = 30$\textdegree, $\phi = 0$\textdegree) under TE polarization, since it is the worst one according to the previous results. The frequency range under examination is normalized to the wavelength of the highest permittivity value (in this case, $\varepsilon_{\text{r}, 3} = 8$). As depicted in Figs.~\ref{S12_dB}-\ref{S12_Phase}, there is a significant agreement between the simulations and the circuital model for both normal and oblique incidence, remarking the accuracy and reliability of our approach. Special attention deserves the smooth variation of the phase when the angle of incidence changes. This \emph{insensitivity} on the phase will be crucial for the design of the TAs. 

\section{Design of fully dielectric TAs for OAM-waves generation}

\begin{figure}[t]
\centering
\subfigure[]{\includegraphics[width=0.48\columnwidth]{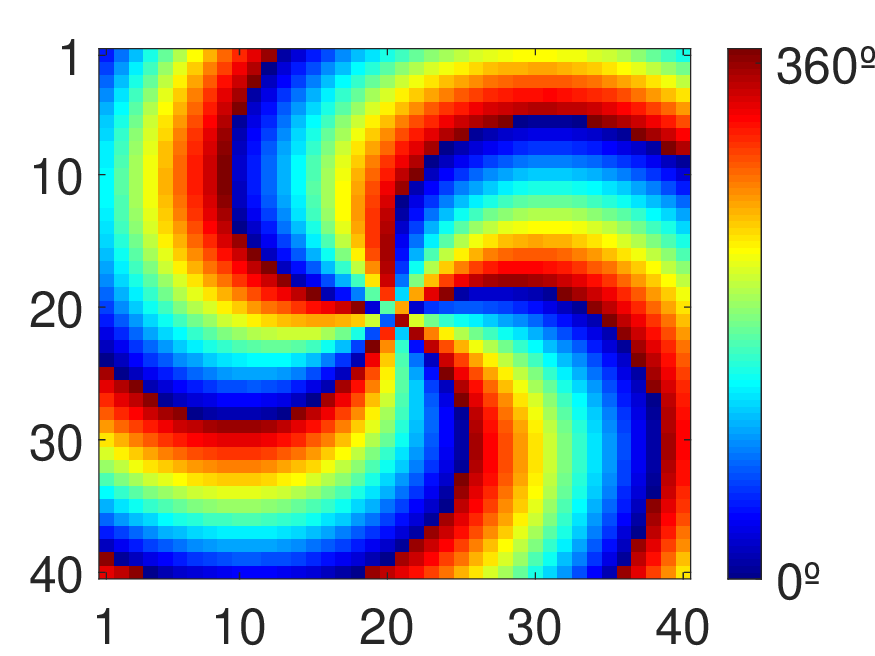}\label{fig:L=5_no_discret}}
\subfigure[]{\includegraphics[width=0.48\columnwidth]{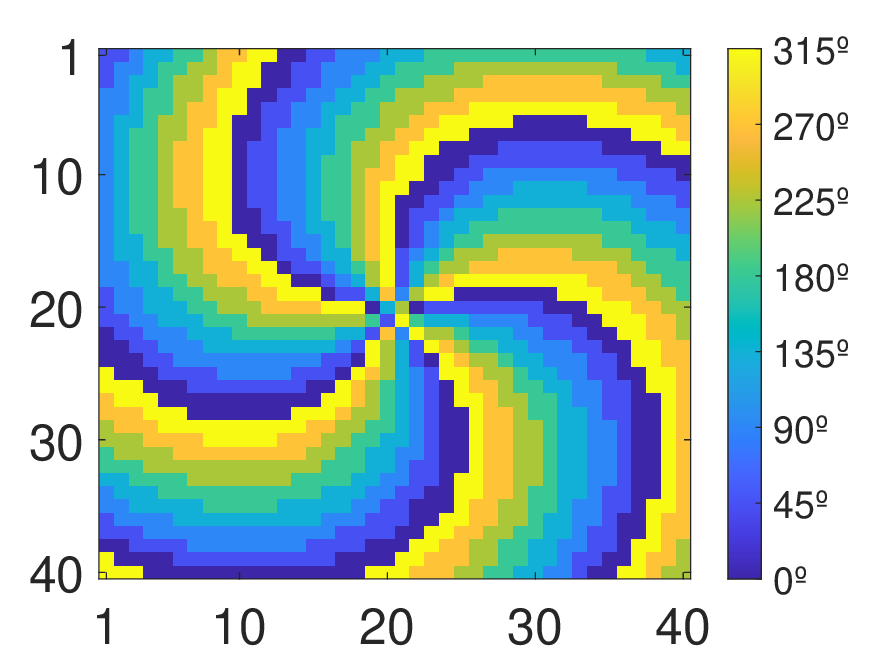}\label{fig:L=5_discret}}    
\subfigure[]{\includegraphics[width=0.48\columnwidth]{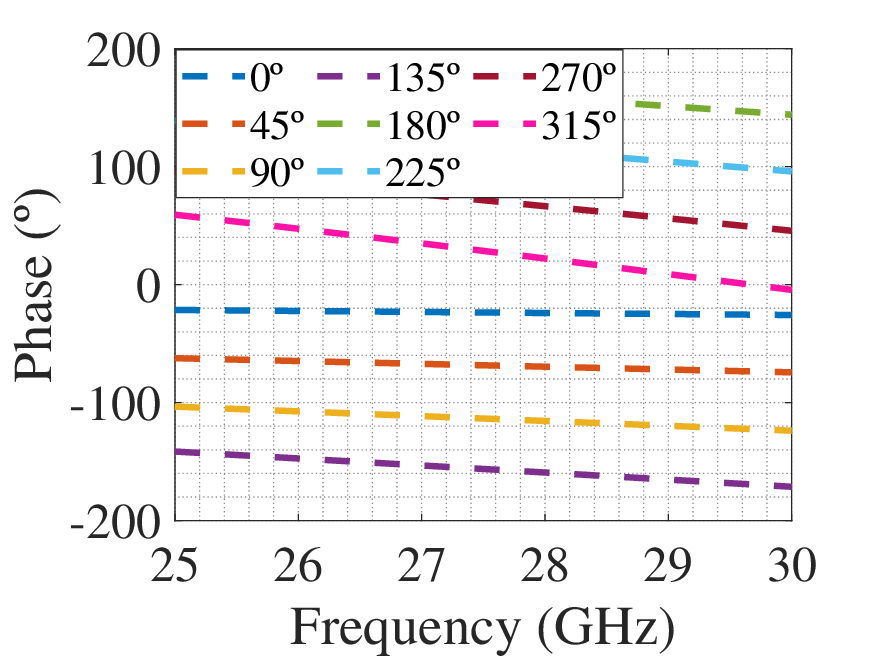}\label{fig:fase_leg}}
\subfigure[]{\includegraphics[width=0.48\columnwidth]{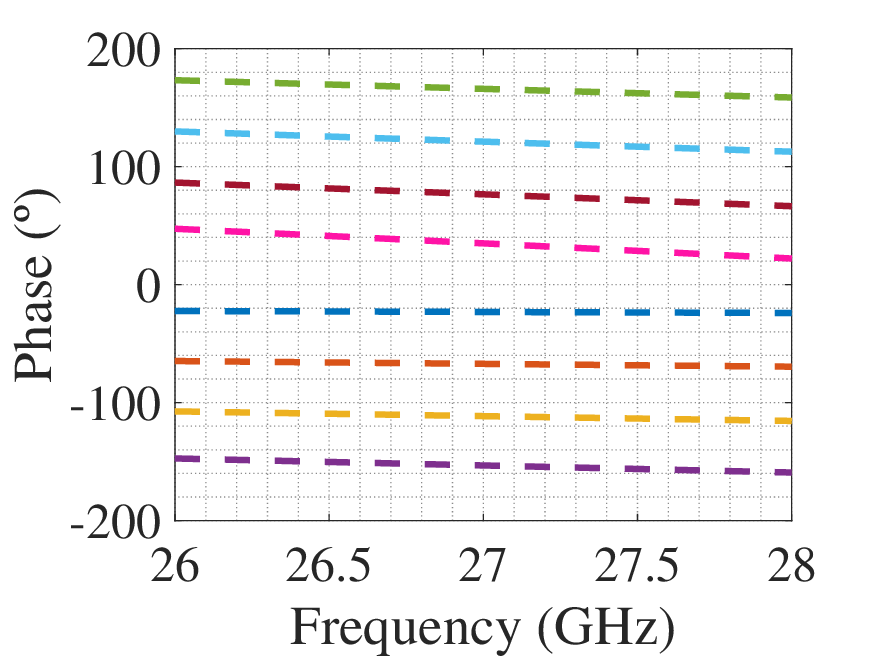}\label{fig:fase_zoom}}
\caption{(a) Phase shift in needed to generate a OAM mode L = 5 (b) Discretization of the phase shift needed to generate a OAM mode L = 5 (c) Phase shift in the 3-bit unit cells (d) Zoom in the band of interest.}
\label{Fig:Discret}
 \end{figure}

For the design of the fully dielectric TAs, the 5G n257 frequency band ($25–30\,$GHz) is selected, since it is one of the promising bands and it is starting to be operative in some regions of the world \cite{Jabbar2025}. An appropriate unit-cell periodicity for this band is $p = 2.7$\ mm, small enough in comparison with the wavelength ($p = 0.25\lambda_0$, being $\lambda_0$ the wavelength in free-space at 28GHz). The total size of the TAs will be conditioned by platforms of $40 \times 40$ unit cells, corresponding to $10.8 \times 10.8\, \text{cm}^{2}$. The fabrication process is carried out via stereolithography through the FormLabs Grey V4 by using a resin having a permittivity of $\varepsilon_{\text{r}} = 2.6$.

\begin{figure}[t]
\centering
\subfigure[]{\includegraphics[width=0.45\columnwidth]{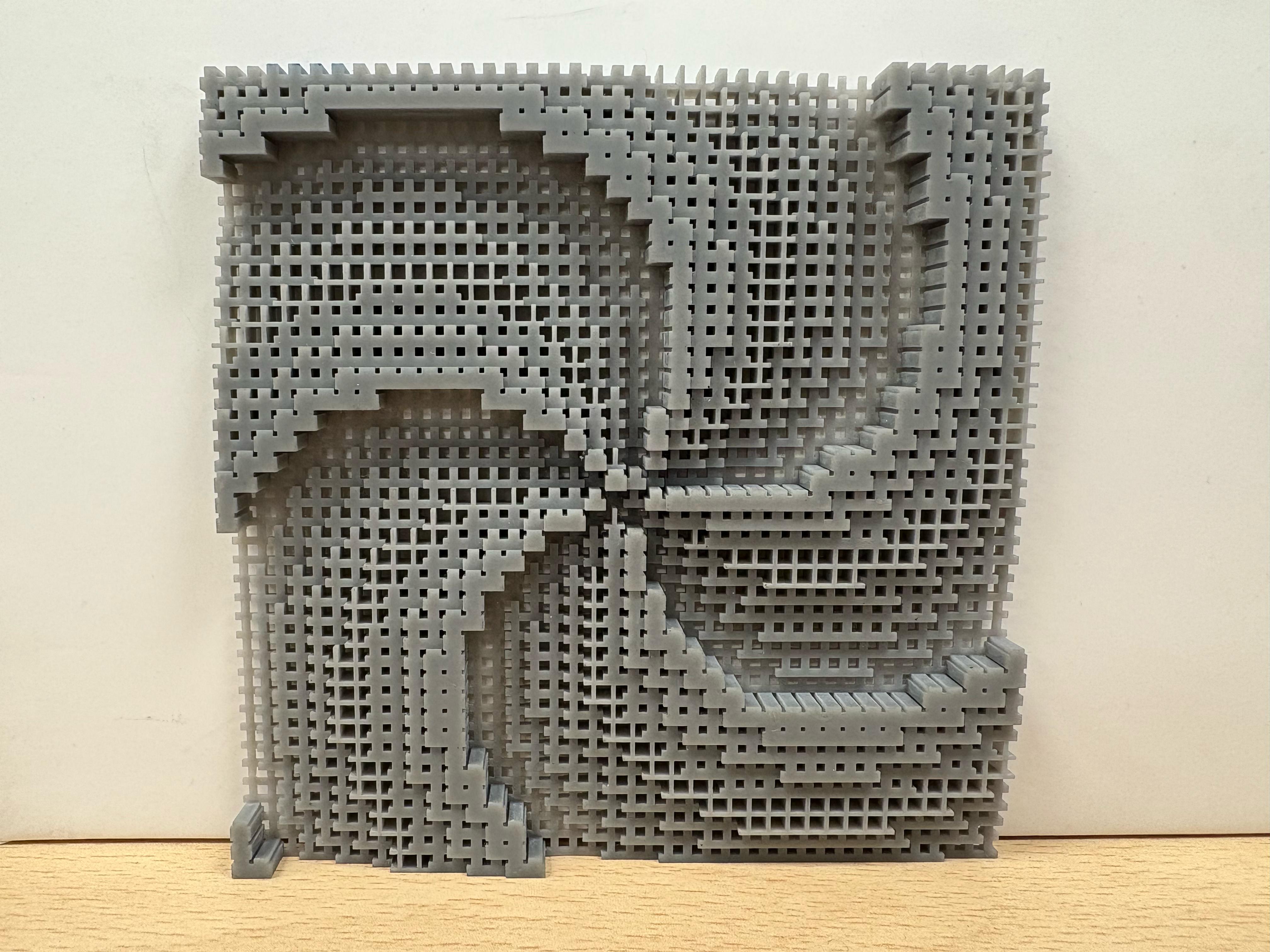}\label{L5_foto}}
\subfigure[]{\includegraphics[width=0.45\columnwidth]{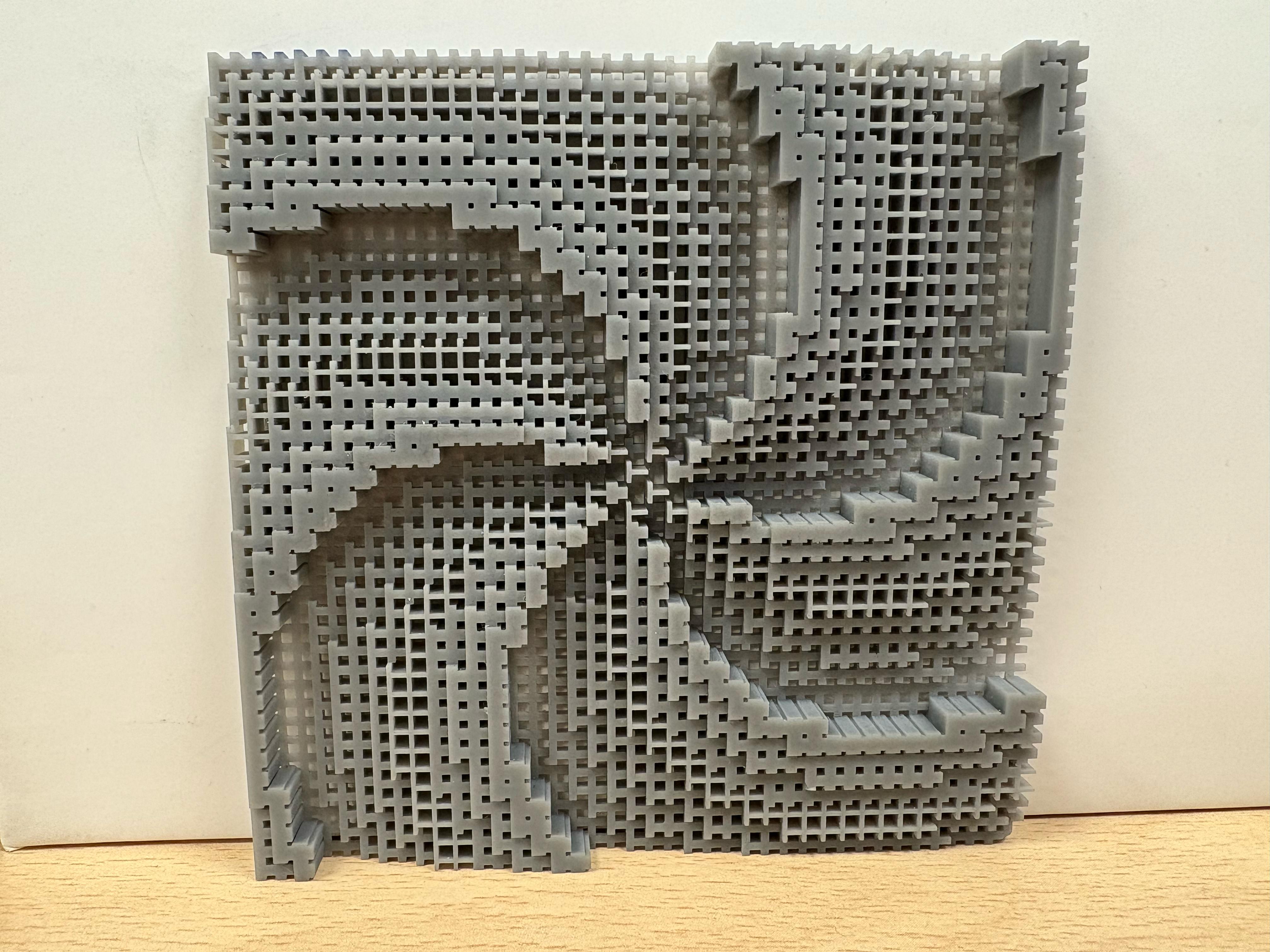}\label{L6_foto}}
\caption{(a) 3D printed TAs to generate OAM with topological charges (orders) (a) L = 5 and (b) L = 6.}
\label{Fig:TA_printed}
\end{figure}

The cell distribution of the TA is designed under the assumption that a spherical wave is impinging upon it. To generate OAMs at the output of the TA, each cell must individually provide a particular phase value $\phi_{m, n}$. This phase value is computed according to:

\begin{equation}\label{eq:OAM}
\phi_{m,n} = L\varphi_{m,n} +k_{0}·|{\Vec{r}_{m,n} -\Vec{r}_f}|
\end{equation}
where
$\Vec{r}_{m,n}$ denotes the position of each unit cell, while $\Vec{r}_{f}$ corresponds to the focal point position. The first term in the equation accounts for the generation of vortex beams, where $L$ is the topological charge (order) of the desired OAM mode \cite{Rahmat2021-OAM}, and $\varphi_{m,n}$ is the azimuthal angle of the $(m,n)$ element. The second term represents the spherical wavefront correction. The summation of both terms provides the phase values in each of the cells. The resulting global output field of the TA comes from the superposition of the individual contributions of the field from each of the cells.  Fig.~\ref{fig:L=5_no_discret} shows an example of the phase distribution (cell by cell) on the TA to generate an output wave with a OAM of order $L = 5$. 

Notice that the phase distribution in Fig.~\ref{fig:L=5_no_discret} has the form of a \emph{continuous} gradient. However, for design simplicity and without any degradation in the performance of the unit cell, the phase gradient is quantized to 3 bits \cite{Nadi2023}. This leads to find 8 phase states, each separated by $45\degree$. The phase distribution under this discretization is shown in Fig.~\ref{fig:L=5_discret}. Each of the states are induced by a different configuration of the unit cell. Fig.~\ref{fig:fase_leg} represents the phase evolution of these states. They have been obtained by the cell configurations in  Table \ref{tab:parametros_3bit}. The Table shows the geometric values associated with the unit cell, along with the $\chi$ that result in the 8 states needed. A plot of the phase shift curves in the band of interest is represented in Figs.~\ref{fig:fase_leg}-\ref{fig:fase_zoom}. 

\begin{table}[h]
\centering
\caption{Parameters of the 3-bit unit cells}
\label{tab:parametros_3bit}
\begin{tabular}{|c|c|c|c|c|c|c|c|c|}
\hline
\textbf{Cell} & \textbf{0º} & \textbf{45º} & \textbf{90º} & \textbf{135º} & \textbf{180º} & \textbf{225º} & \textbf{270º} & \textbf{315º} \\ \hline
\textit{\textbf{$w (mm)$}} & 1.0  & 1.0  & 1.1  & 1.0  & 0.5  & 1.4  & 1.4  & 1.9  \\ \hline
\textit{\textbf{$d (mm)$}} & 0.5  & 1.5  & 2.5  & 3.5  & 5.0  & 5.0  & 6.0  & 6.5  \\ \hline
\textbf{$\chi$}            & 0.60 & 0.60 & 0.65 & 0.60 & 0.34 & 0.77 & 0.77 & 0.91 \\ \hline
\end{tabular}
\end{table}

In order to account for the experimental tests in the near field, and the influence on the topological charge $L$ in the sucess/failure on the reception of information, 3 TAs with different topological charges have been fabricated: $L = 3, 5, 6$. Notice that these TAs are also valid to induce OAMs with the opposite topological charge $L = -3, -5, -6$ when they are fed from the opposite side. Two of these prototypes are shown in Fig.~\ref{Fig:TA_printed}, consisting of TAS designed to generate OAM-waves with $L = 5$ and $L = 6$ respectively.   

\begin{figure}[b]
\centering    
\subfigure{\includegraphics[width=0.40\columnwidth]{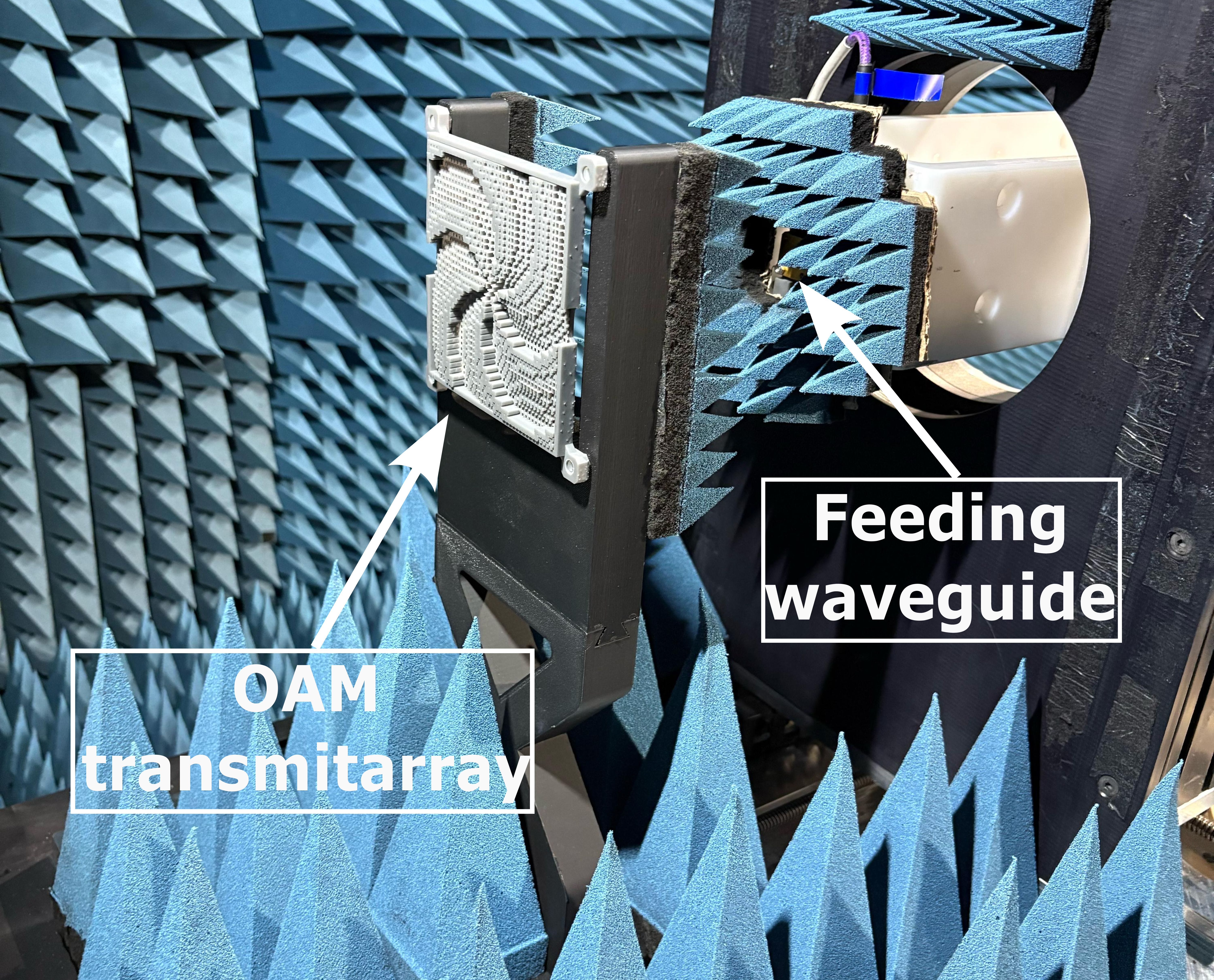}\label{fig:Setup_abs}}
\subfigure{\includegraphics[width=0.49\columnwidth]{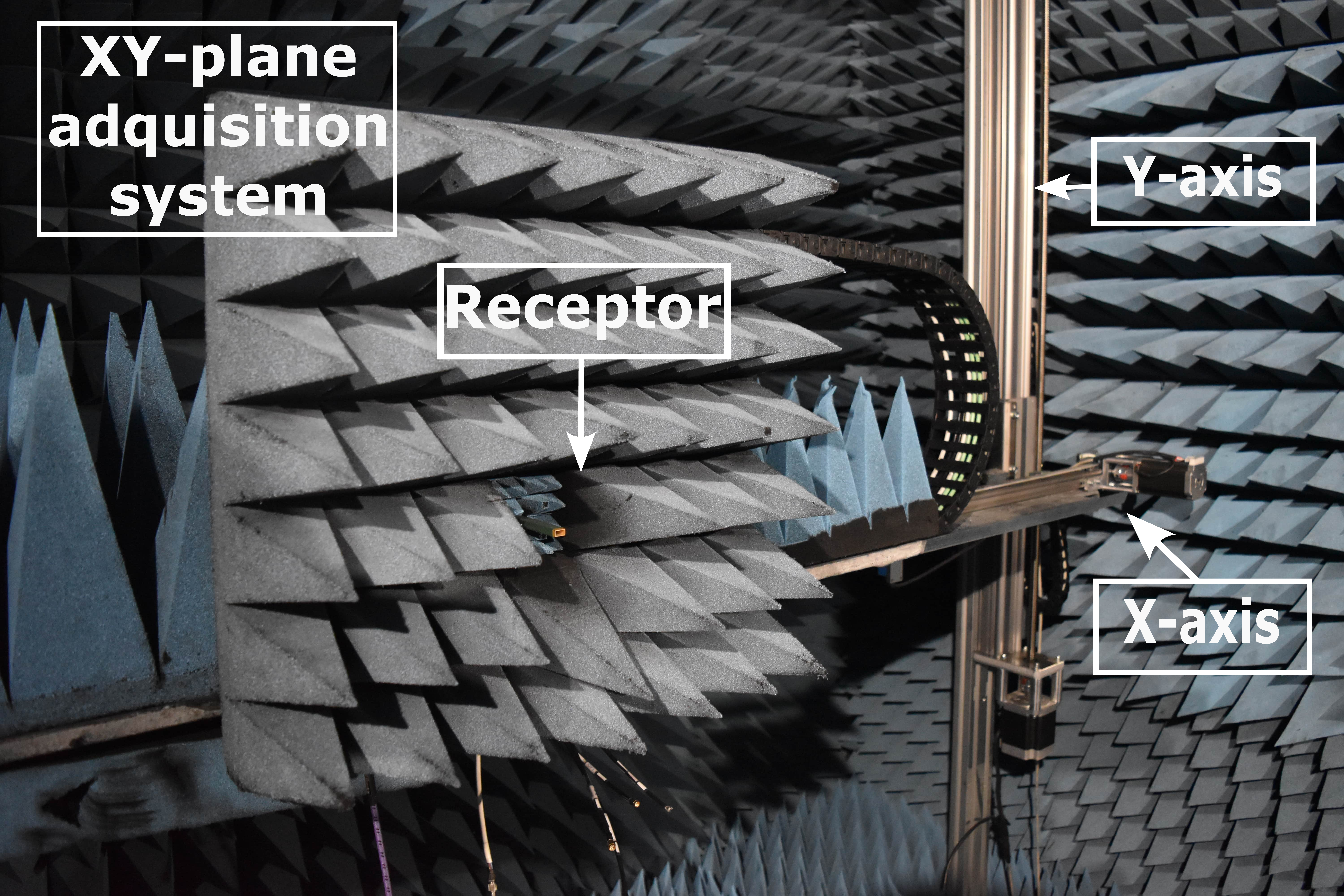}\label{fig:XY_system}}
\subfigure[]{\includegraphics[width=0.49\columnwidth]{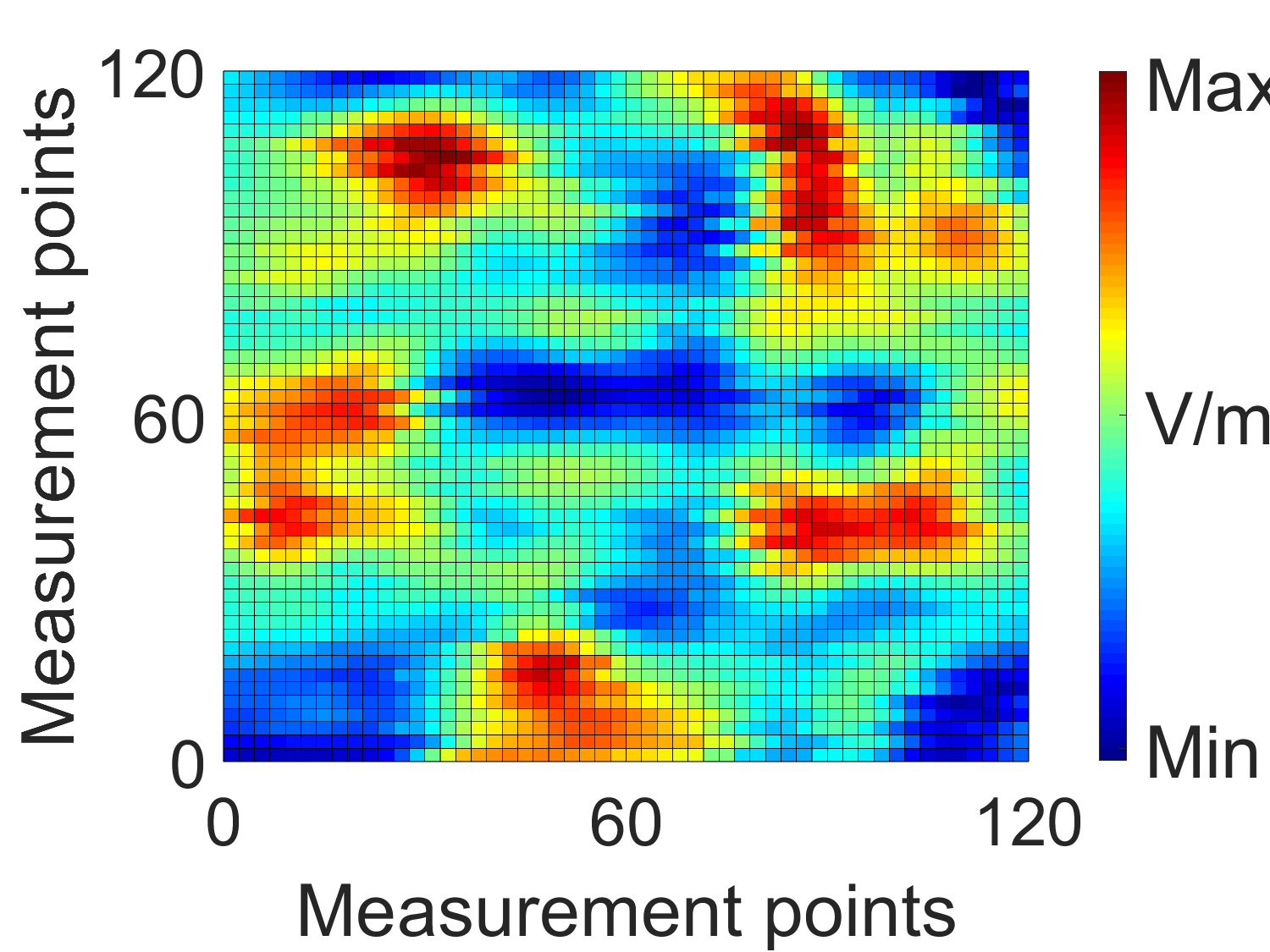}\label{fig:L5_lab}}
\subfigure[]{\includegraphics[width=0.49\columnwidth]{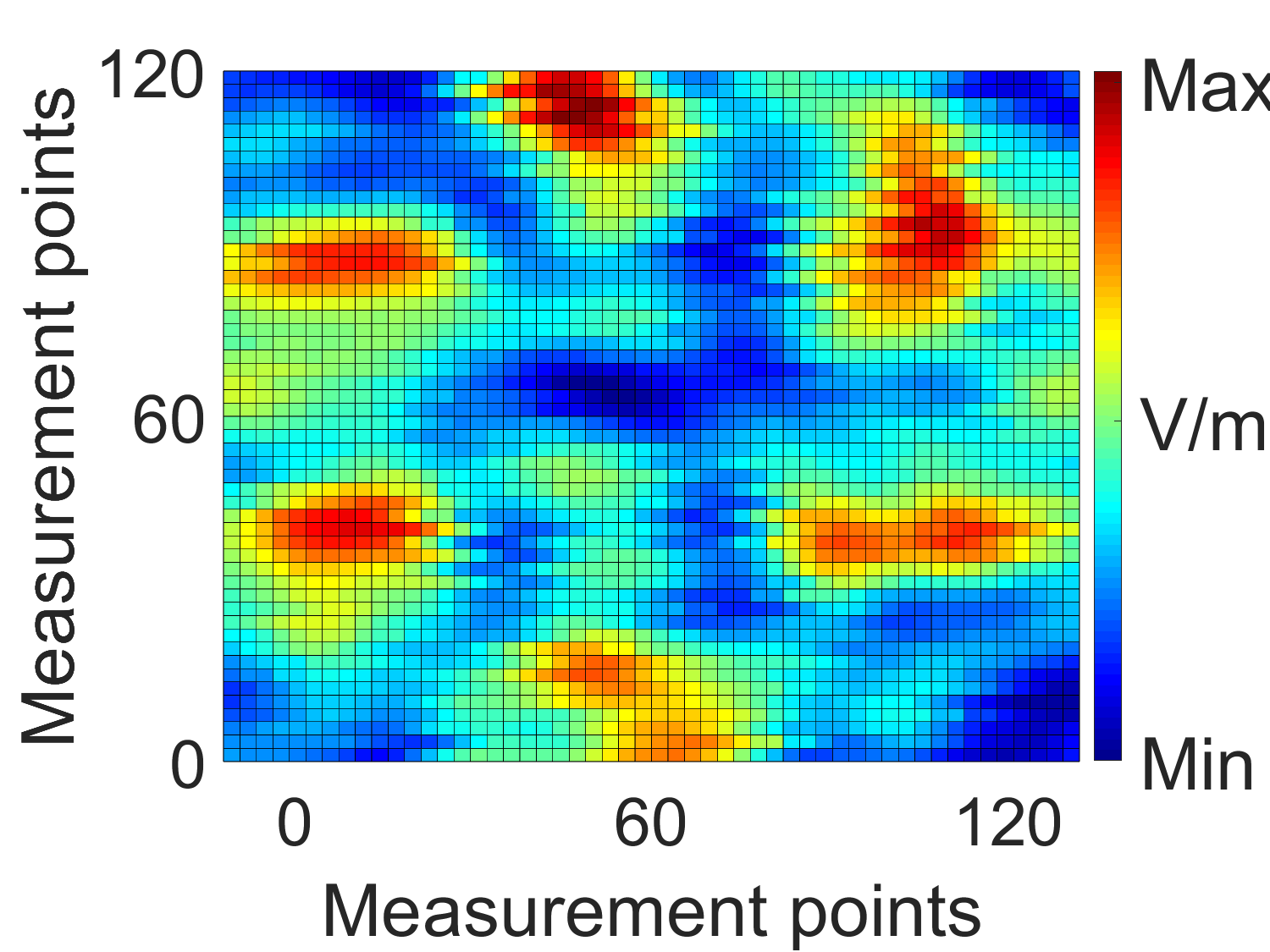}\label{fig:L6_lab}}
\caption{(a) Transmitting system including the feeder and the TA. (b) Receiving setup formed by a planar scanning system. (c) E-field amplitude pattern at 28 GHz for the mode $L=5$. (d) Same for the mode $L=6$. The axis denote the number of pixels regarded during the measurement process.}
\label{Fig:OAM_lab}
 \end{figure}

\section{Validation of the TAs and experimental tests for communication links}

The experimental validation of the TAs emulating near-field communication scenarios  
were conducted at the Smart Wireless Technology laboratory at University of Granada (UGR) \cite{SWAT}. The TAs are fed by WR34 open-ended waveguides, whose operation band goes from 22 to 33 GHz. Each waveguide provides a gain of 8.1\:dBi and exhibits half-power beamwidths (HPBWs) of 30.0º in the H-plane and 37.6º in the E-plane. They are placed at a distance of $|\Vec{r}_{f}| = 20\,$cm from the TAs. For the experimental scenario, they emulate efficiently a focal-point source thus the radiated electromagnetic fields are approximately regarded as spherical waves. 

\begin{figure}[]
\centering
\subfigure[]{\includegraphics[width=0.345\columnwidth]{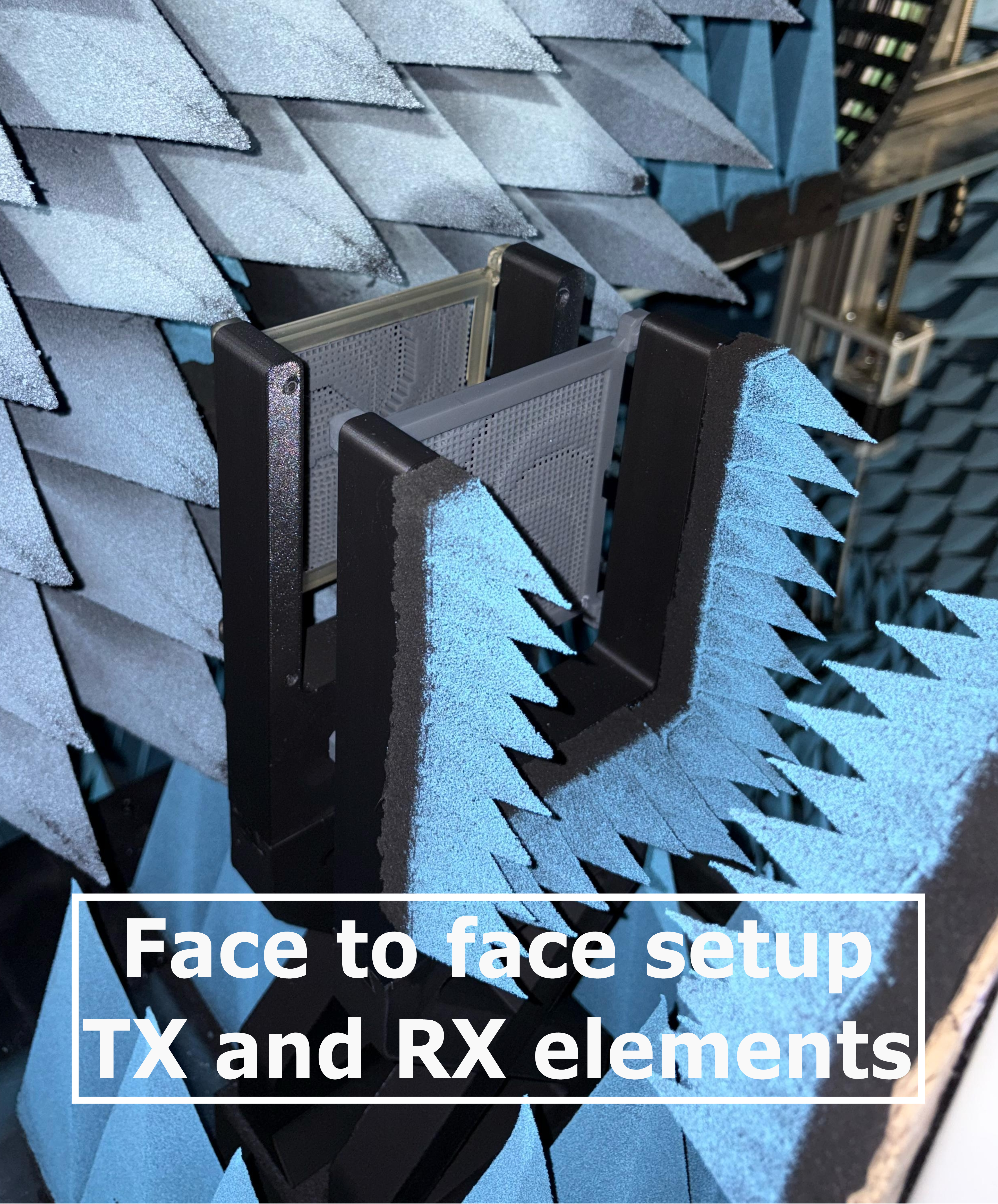}\label{fig:enfrentados}}
\subfigure[]{\includegraphics[width=0.55\columnwidth]{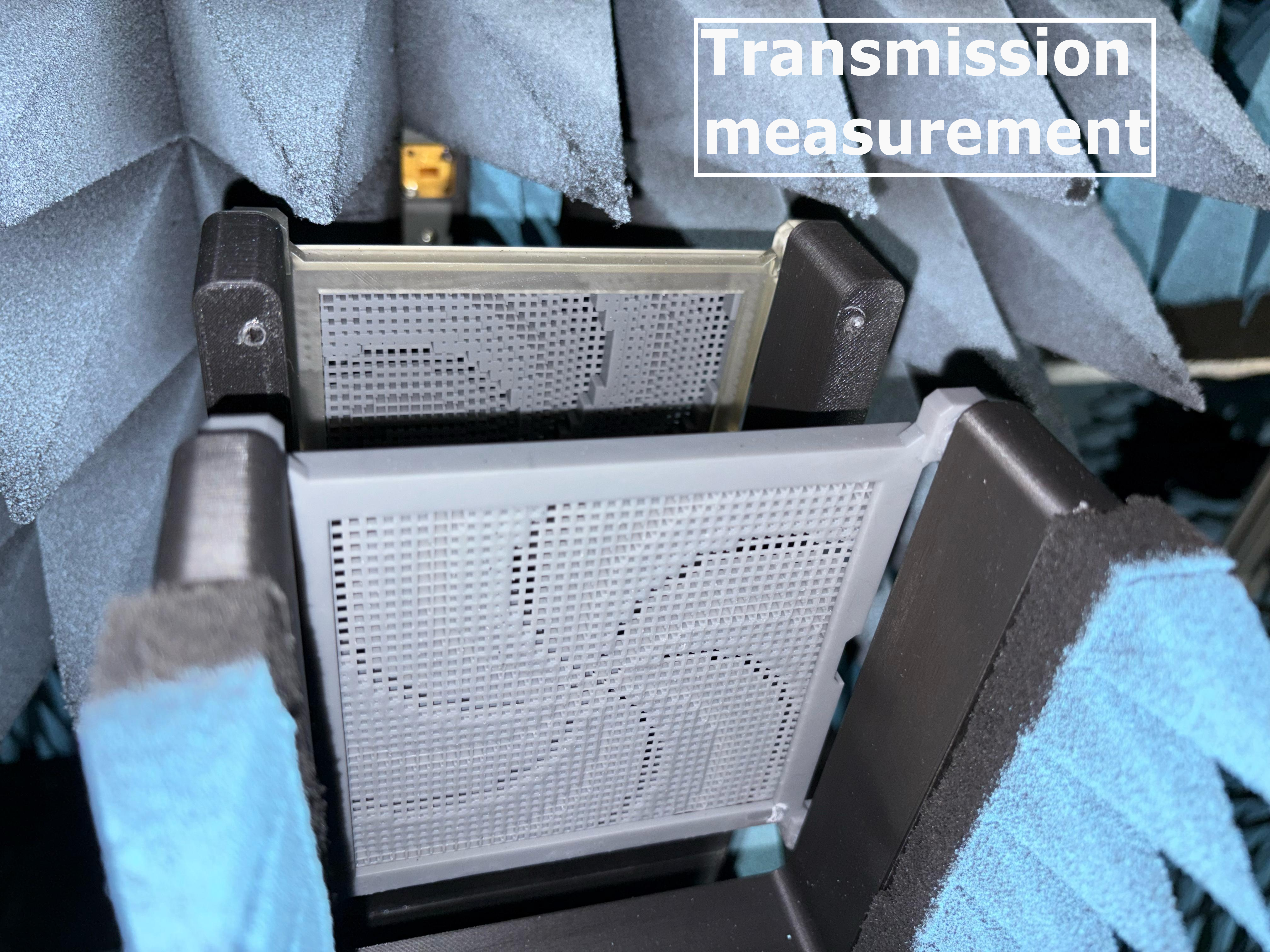}\label{fig:enfrentados2}}
\subfigure[]{\includegraphics[width=0.9\columnwidth]{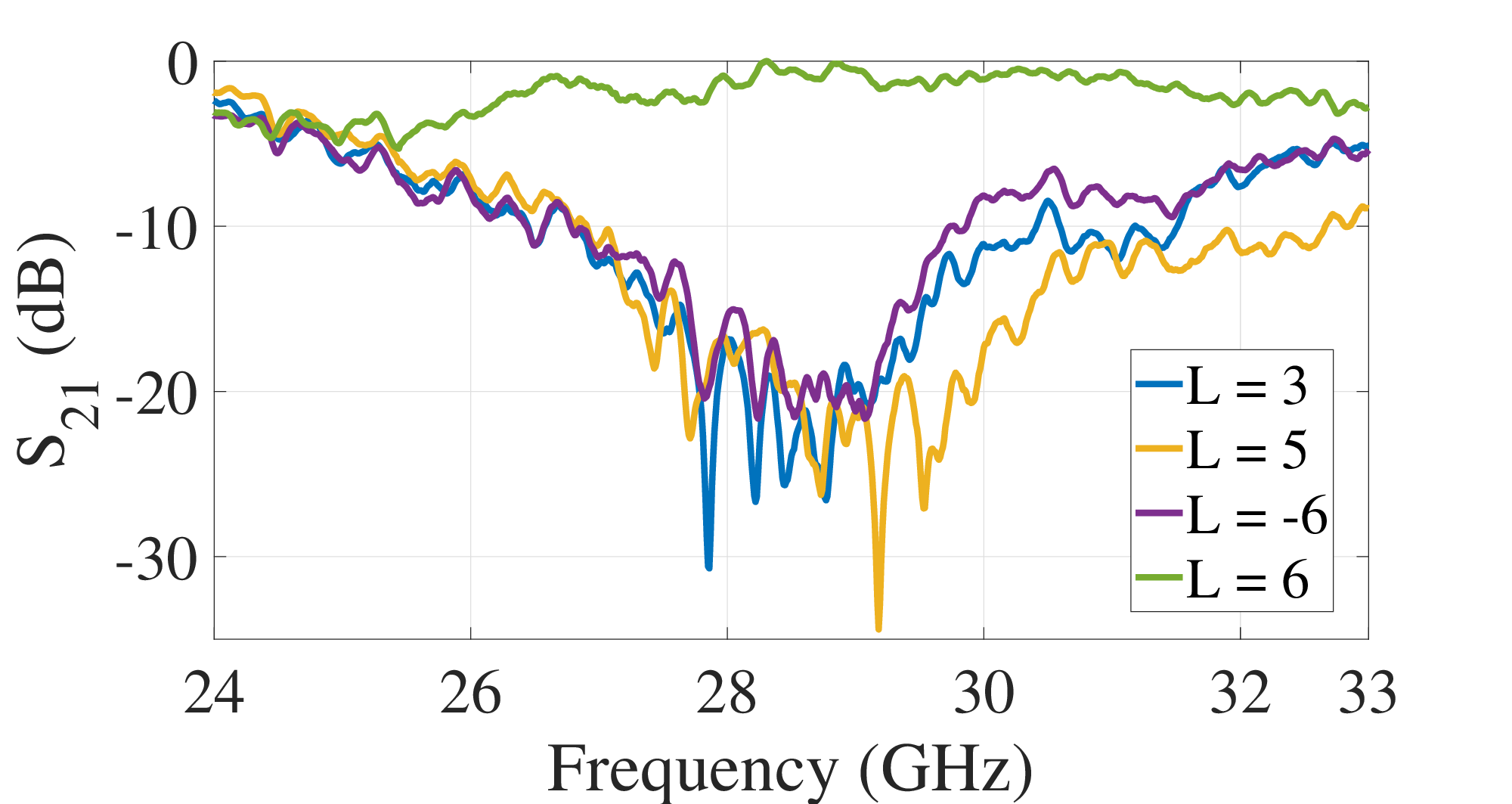}\label{fig:OAM_L6_norm_7.5}}
\subfigure[]{\includegraphics[width=0.9\columnwidth]{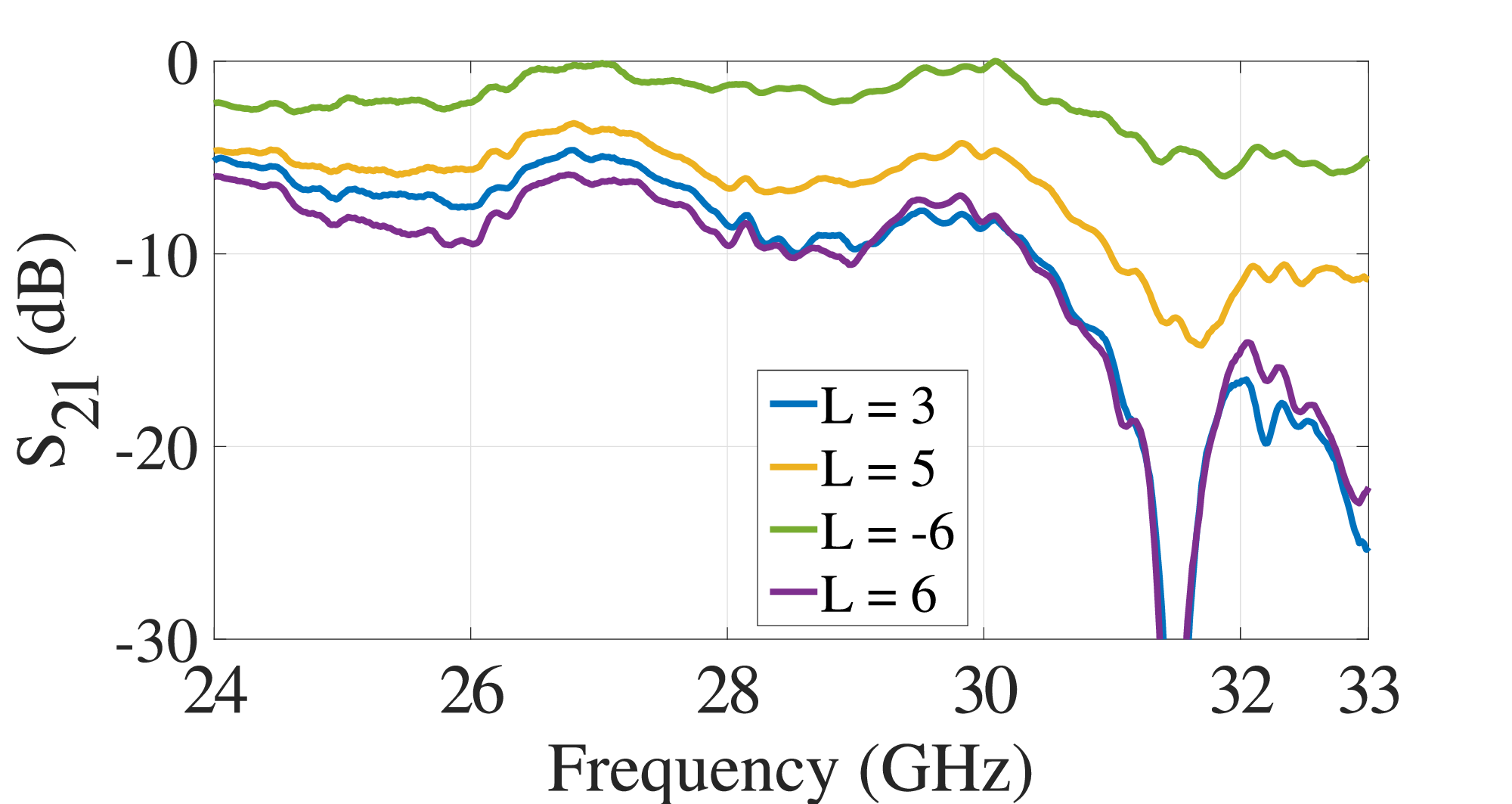}\label{fig:OAM_L6_norm_8.5}}
\caption{Design measurement setup: (a) Supports to place the TAs (b) System with two confronted TAs. Measurements to evaluate the orthogonality of the modes, L=6 is the transmitter. (c) Distance between the TAs is 7.5cm (d) Distance between the TAs of 8.5cm.}
\label{Fig:Setup_completo}
\end{figure}

\subsection{Characterization of the TAs}
A first measurement campaign consisted on the characterization of the vortex-beam patterns generated by the TAs. Figs.~\ref{L5_foto}-\ref{L6_foto} show two paradigmatic examples of TAs, to generate both vortex beams with $L = 5$ and $L = 6$ respectively. The corresponding experimental setup is shown in Figs.~\ref{fig:Setup_abs}-\ref{fig:XY_system}, exhibiting the TX system formed by the feeding waveguide plus the TA; and the RX system consisting of a scanning system in the XY-plane. The distance between the TX systems (from the TA) and the RX system is set at $20\,$cm. The planar scanning system is formed by a moving probe sweeping the XY plane in steps of $\lambda_{0}/4$. The probe recovers the field at each spatial step with the aim of elaborating a field map along the whole acquisition plane. Figs.~\ref{fig:L5_lab}-\ref{fig:L6_lab} represent the electric-field map obtained by the TAs in Fig.~\ref{Fig:TA_printed} at 28 GHz. The axis of the figures denote the number of pixels employed in the measurement process, labeled as measurement points in the figures. Since the minimum step is $\lambda_{0}/4 \approx 2.9\,$mm, the acquisition plane takes $35\times35\,\text{cm}^2$ approximately, needing a total of 120 points. The inherent vorticity and the corresponding number of lobes are clearly observable in both cases, corresponding with their OAM-order.

\subsection{Experimental validation of OAM orthogonality}

Once the vortex radiation patterns are fully characterized, a second campaign of measurements is conducted to evaluate the orthogonality between different TAs. This experimental test is conceived to evaluate a scenario where the information is coded in terms of different OAM modes. The idea is to emulate a communication system formed by two reconfigurable TAs: one operating as TX and one as RX. The ideal reconfigurability would allow to re-adapt the phase distribution of the TAs to transmit/receive a particular OAM mode. Since our TAs are not reconfigurable, we proceed by interchanging different static TAs and evaluating the quality of the reception in terms of the $S_{21}$-parameter. The corresponding measurement setup is shown in Figs.~\ref{fig:enfrentados}-\ref{fig:enfrentados2}. As can be observed, two individual TAs are placed face-to-face and precisely aligned. Both TAs are conveniently fed by two independent and well isolated WR34 feeders. Each of the TAs are mounted on a support, which in turn are attached to a rail system that allows for mobility. Thanks to this, the distance between the TAs can accurately be adjusted. The setup is prepared to easily place/remove different TAs.

The measurements were conducted for 2 different distances between the TX/RX, specifically for $7.5\,$cm and $8.5\,$cm. These distances are of the order of the TAs size, thus guaranteeing that the OAM divergence is not substantial. Notice that we can play with the size of the TAs in order to increase/decrease the link distances. For both specific cases, the TX system is configured to generate an OAM mode with order $L = 6$. In the first case, assuming a distance of $7.5\,$cm, the RX system is configured to receive OAMs with orders $L = 3, 5, 6, -6$. Ideally, good reception is expected when TX and RX share the same OAM order. The receiving power, otherwise, must degrade when these orders do not coincide due to the inherent orthogonality. Fig.~\ref{fig:OAM_L6_norm_7.5} shows the $S_{21}$ parameter for all the cases. The results have been normalized to the $S_{21}$-parameter associated with the case where the TX and RX share orders ($L = 6$). The normalization is useful to evaluate and compare different situations. As expected, it can be seen a drop on the reception amplitude when the TX and RX orders are different from each other. It is interested to note that at the design frequency, $28\,$GHz, the difference between the reception power when the orders are matched and not matched is about $20\,$dB. This difference reduces as we move from the design frequency. In fact the communication system is efficient from $27$ to $30\,$GHz approximately (relative bandwidth of about $10\%$), where a threshold of $10\,$dB between the matched and mismatched cases is established. A \emph{secure} communication link is established along this operation band. The information is coded by a given $L$-order, and can be decoded if the RX system is configured with the same order.     

Fig.~\ref{fig:OAM_L6_norm_8.5} shows the results obtained in a similar scenario, but now fixing a distance of $8.5\,$cm between TX/RX. Notice that the effects related to the orthogonality in the communication system are maintained, but the reception quality has decreased for the matched case. This is an effect of the OAM-diverging nature. In general, since OAMs are relevant in near-field contexts, the distance between TAs is a critical parameter. Good reception is achieved when the phase distributions in the TX/RX transmitarrays are configured by previously taking into account the distance between them. This finds similarities with systems formed by lenses, where the focus distance is crucial \cite{Torcolacci2023}. In the spatial region existing between TX and RX when the OAM is perfectly focused, the OAM evolution is convergent instead of divergent, and most of the original power can be captured. Future reconfigurable and intelligent systems must not only support reliable reconfiguration capabilities, but also enable accurate estimation of the target or user distance in order to ensure an efficient communication link.

\subsection{Self-healing and object avoidance}

\begin{figure}[t!]
\centering   
\subfigure[]{\includegraphics[width=0.41\columnwidth]{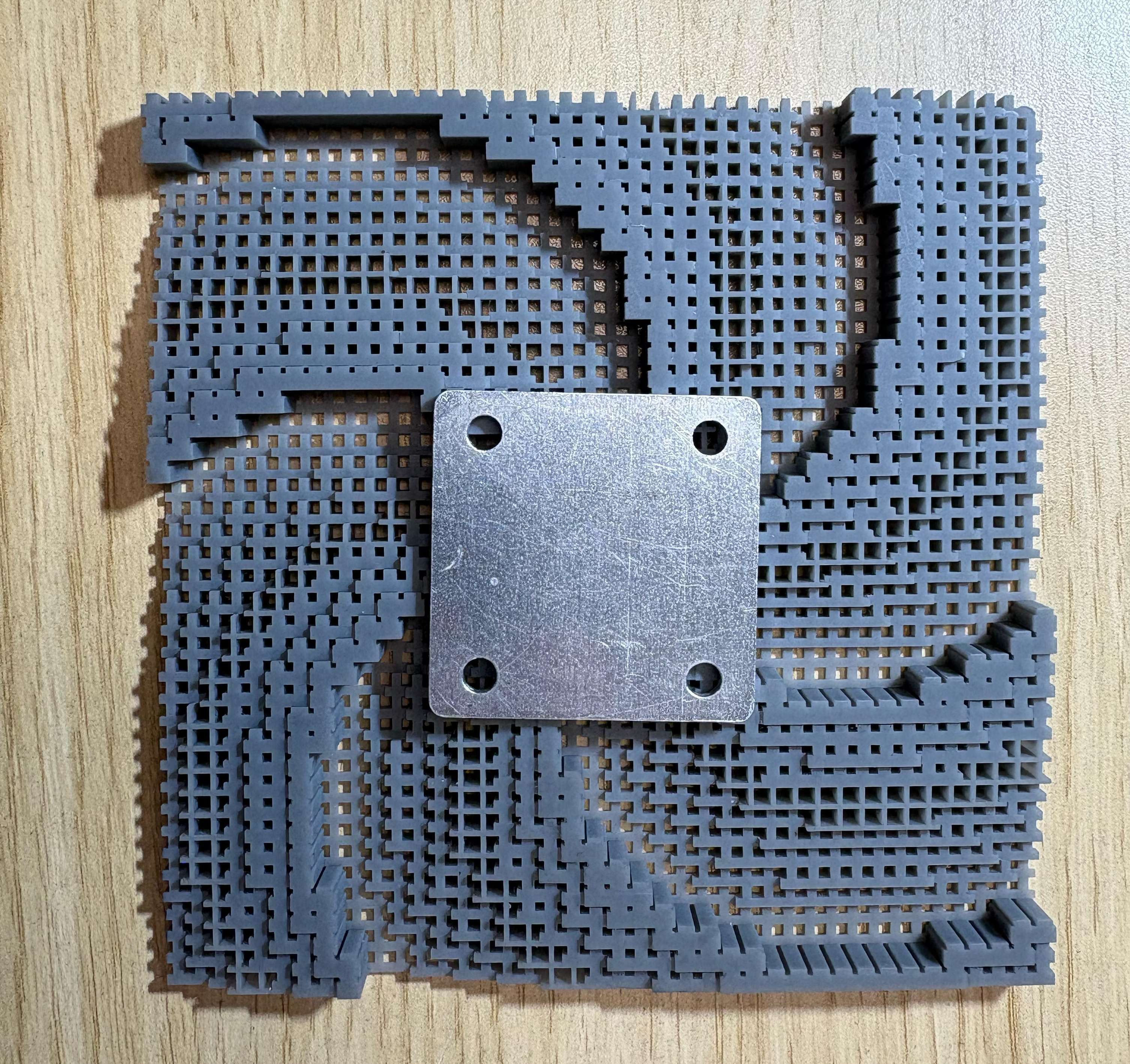}\label{fig:WR75}}
\subfigure[]{\includegraphics[width=0.55\columnwidth]{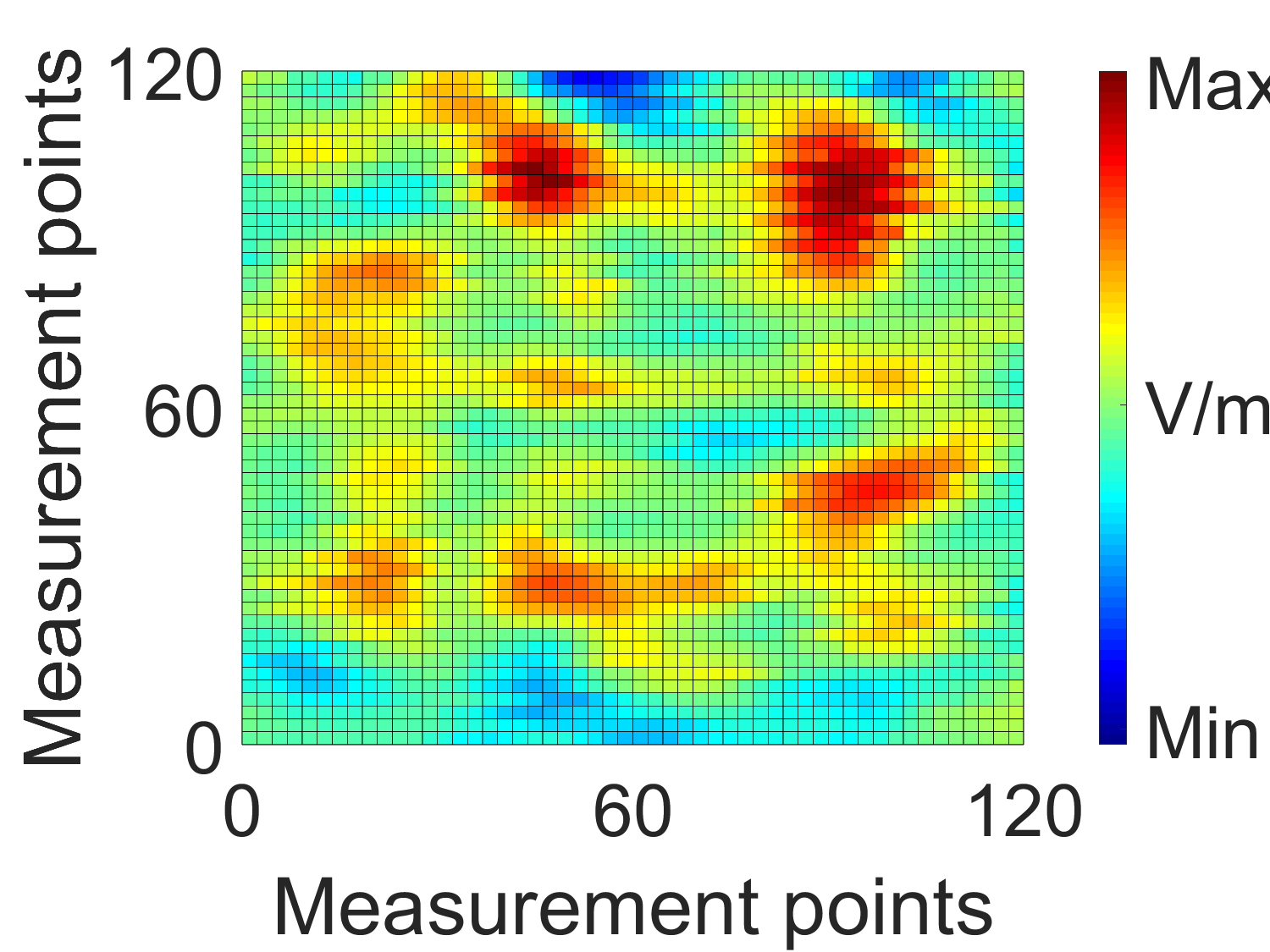}\label{fig:L6_28Ghz_20cent}}
\subfigure[]{\includegraphics[width=0.9\columnwidth]{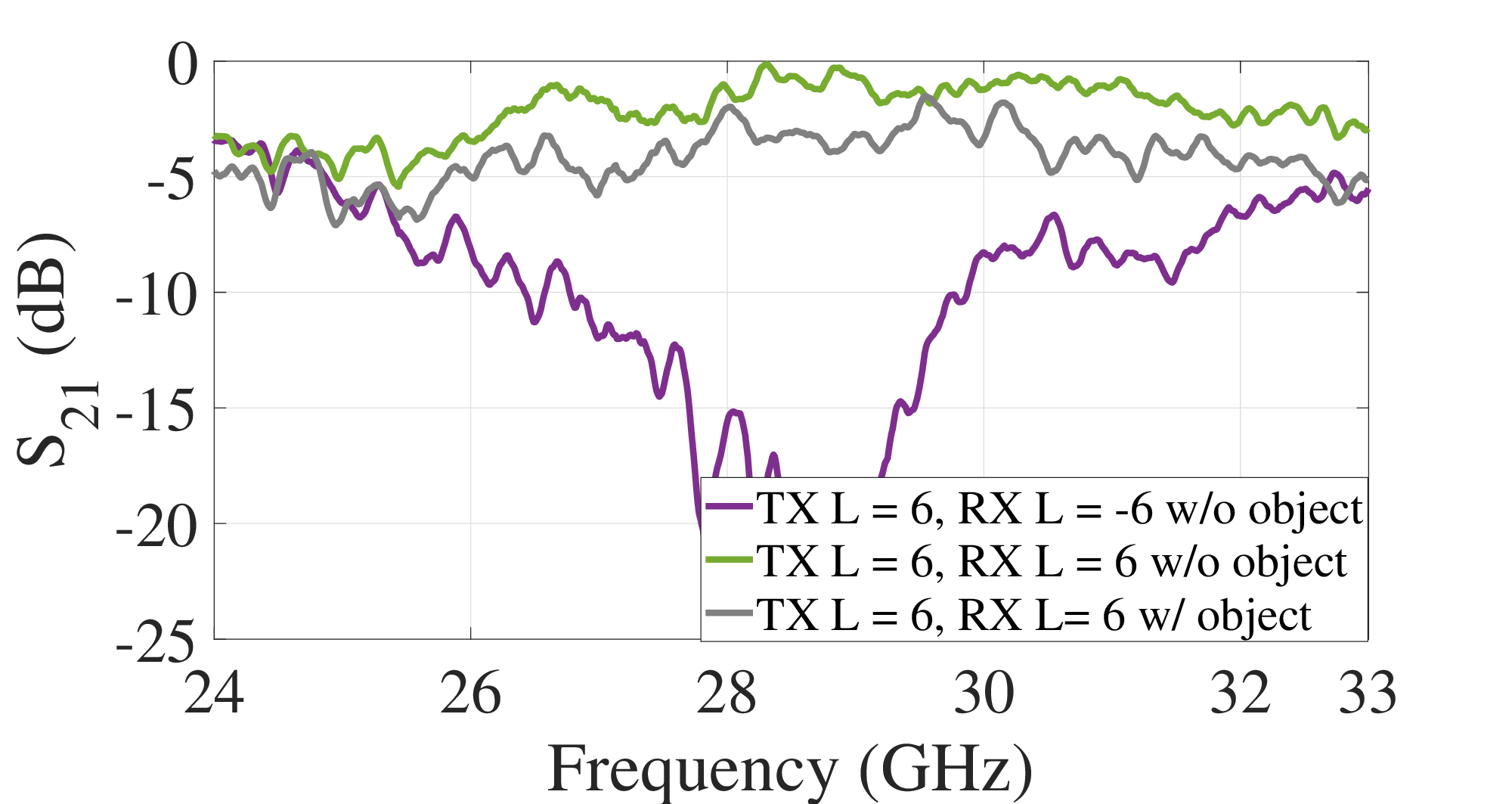}\label{fig:OAM_objects}}
\caption{(a) Picture of the metallic object over the TA with $L=6$. (b) Characterization of the mode with order $L=6$ at $28\,$GHz after overpassing the object. (c) $S_{21}$ measurement with a $L=6$ transmitter and a receptor $L=6$ and $L=-6$. The distance between them is the $7.5\,$cm. When the transmitter and the receptor match, it is measured with and without an object at a intermediate distance.}
\end{figure} 

A final experimental test is conceived to evaluate the capacity of the OAM waves to \emph{avoid} objects and the capability of self-reconstruction. A square-shaped metal with dimensions $38.45 \times38.45\,$mm$^{2}$ is placed in the intermediate distance between both TAs. Fig.~\ref{fig:WR75} shows the object size in comparison to the size of the TA. As can be appreciated, this size is non-negligible, occupying a surface of about 8-9 times smaller than the surface of the TAs. 

The organization of the experiment is the following: the OAM is again characterized as was done in Figs.~\ref{fig:L5_lab}-\ref{fig:L6_lab}, but assuming the presence of the square-shape metal at a distance of $3.75\,$cm from the TA. The field obtained is plotted in Fig.~\ref{fig:L6_28Ghz_20cent}, where we have assumed the TA with order $L = 6$. By comparing this field with that characterized in Fig.~\ref{fig:L6_lab} (no object in the pathway), we can see how the vorticity is maintained despite a slight deterioration. This result is consistent with previous assumptions and confirms the object-avoiding or self-reconstruction capacity of OAM waves. And of course, this suggests that the link between TX and RX with same orders could still be operative. This is corroborated by using the setup described in Figs.~\ref{fig:enfrentados}-\ref{fig:enfrentados2}. The distance between TAs is set at $7.5\,$cm, which has been proven to be an optimal distance. The object is placed at the intermediate distance, that is, at $3.75\,$cm. Again, the $S_{21}$ parameter is used to evaluate the reception power, as plotted in Fig.~\ref{fig:OAM_objects}. For the sake of a fair comparison, the figure includes results obtained when no object exists in the spatial region between the TAs for the matched case ($L = 6$ in both TAs) and for the mismatched case ($L = 6$ in TX, $L = -6$ in RX). Again, the results have been normalized to the optimal case. As can be observed, the received power when the object is included in the communication link is slightly degraded, but the communication is maintained. The power difference between the cases with and without object ranges from $2$–$3\,$dB, in contrast to the difference of up to $20\,$dB observed when the receiving order is $L = -6$. This indicates that the non-orthogonality has a deeper impact on the degradation of the communication than the present of objects, and enables these platforms as future key communication systems for indoor and relatively crowded environments for the 5G era.   

\begin{table}[t]
\centering
\caption{Parameters of the 3-bit unit cells for OAMs communication}
\label{tab:parametros_3bit_comms}
\begin{tabular}{|c|c|c|c|c|c|c|c|c|}
\hline
\textbf{Cell} & \textbf{0º} & \textbf{45º} & \textbf{90º} & \textbf{135º} & \textbf{180º} & \textbf{225º} & \textbf{270º} & \textbf{315º} \\ \hline
\textit{\textbf{$w (mm)$}} & 1.2  & 1.0  & 1.1  & 1.1  & 1.0  & 1.1  & 1.0  & 1.0  \\ \hline
\textit{\textbf{$l (mm)$}} & 24.4  & 24.3  & 19.1  & 15.7  & 13.0  & 8.7  & 5.7  & 2.0  \\ \hline
\textbf{$\chi$}            & 0.69 & 0.60 & 0.65 & 0.65 & 0.60 & 0.65 & 0.60 & 0.60 \\ \hline
\end{tabular}
\end{table}

\section{Field-test of OAM-based communications}
In the previous sections, an extensive characterization and validation of the OAM design has been performed within the anechoic chamber. Upon this point, the prototype is ready to be tested under real conditions within a short-range test bench as part of a wireless communications system with minor changes. In particular, the parameters of the unit cell [see Fig.~\ref{T_perfil}] have been slightly modified to accurately focus the OAM wave at a fixed distance. These can be consulted in Table \ref{tab:parametros_3bit_comms}, and it is worth noting that periodicity remains $p$\:$=$\:2.7\:mm as in the other configurations. 

The experimental setup consisted on two OAM transmitarrays aligned in line-of-sight (LoS) and designed for a frequency of 28\:GHz. These two transmitarrays were fed with open-ended rectangular waveguides WR34, both being the same as the employed in the previous section. The focal distance between the source feed and the transmitarray was fixed to 20\:cm. On the other hand, two different pairs of transmitarrays were manufactured to produce $L$\:$=$\:2 and $L$\:$=$\:6 with focus distances for the OAMs of 30\:cm and 14\:cm respectively. Measurements were conducted using a vector signal generator at the TX side and a signal and spectrum analyzer at the RX side. The models are SMW200A and FSW67 from Rohde \& Schwarz\textsuperscript{\textregistered}, both capable of transmitting/receiving signals up to 67\:GHz. The transmitted signal was modulated with QPSK scheme at a symbol rate of 10\:Msym/s, and the bit streams were obtained from a pseudorandom binary sequence PRBS7. At the receiver side, 100 realizations of 1\:ms signals were acquired with 100\:MHz of sample rate.

\begin{figure}[t!]
\centering    
\subfigure[]{\includegraphics[width=0.75\columnwidth]{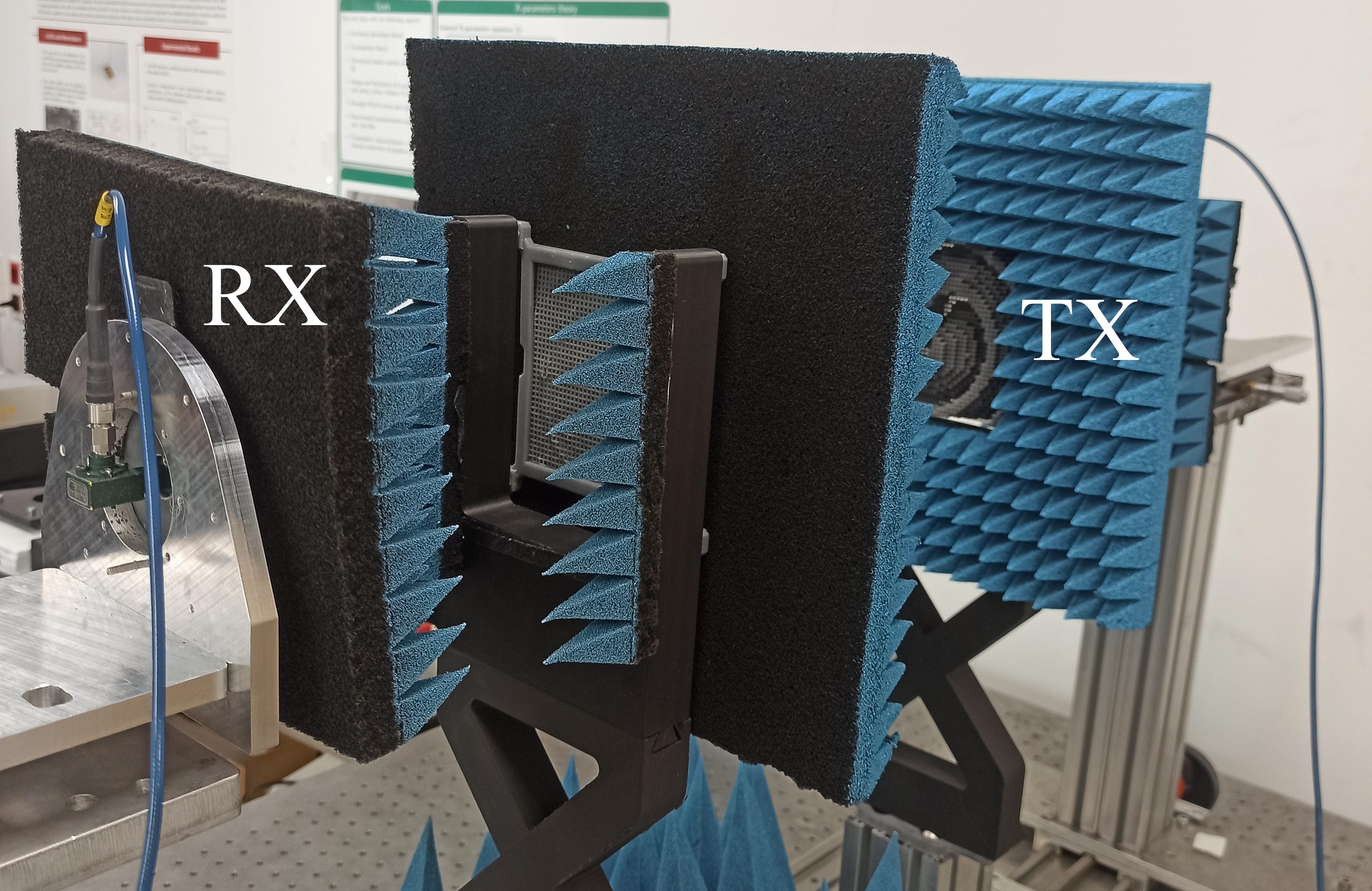}\label{fig:Setup_comms}}
\subfigure[]{\includegraphics[width=0.49\columnwidth]{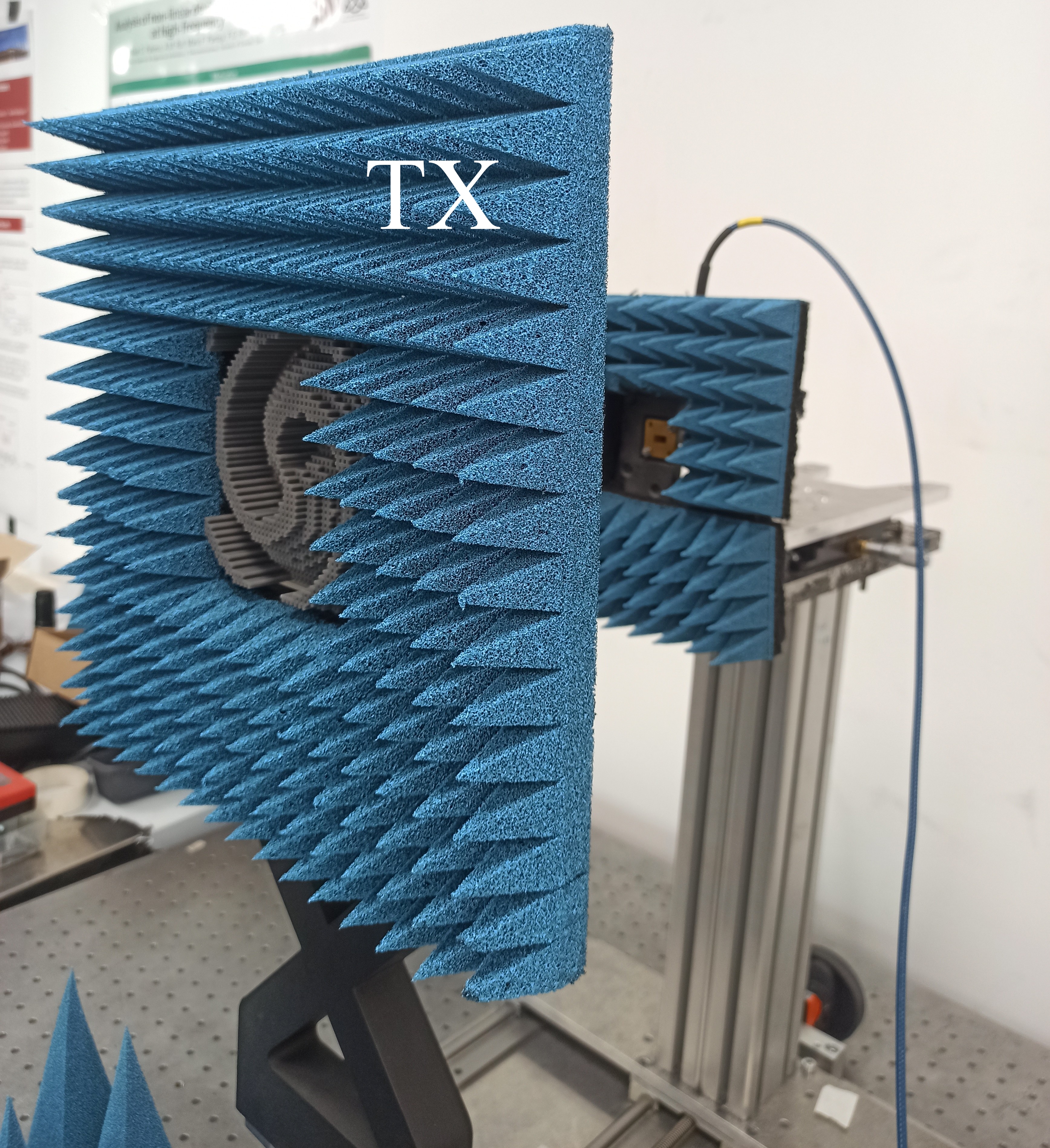}\label{fig:emitter_comms}}
\subfigure[]{\includegraphics[width=0.49\columnwidth]{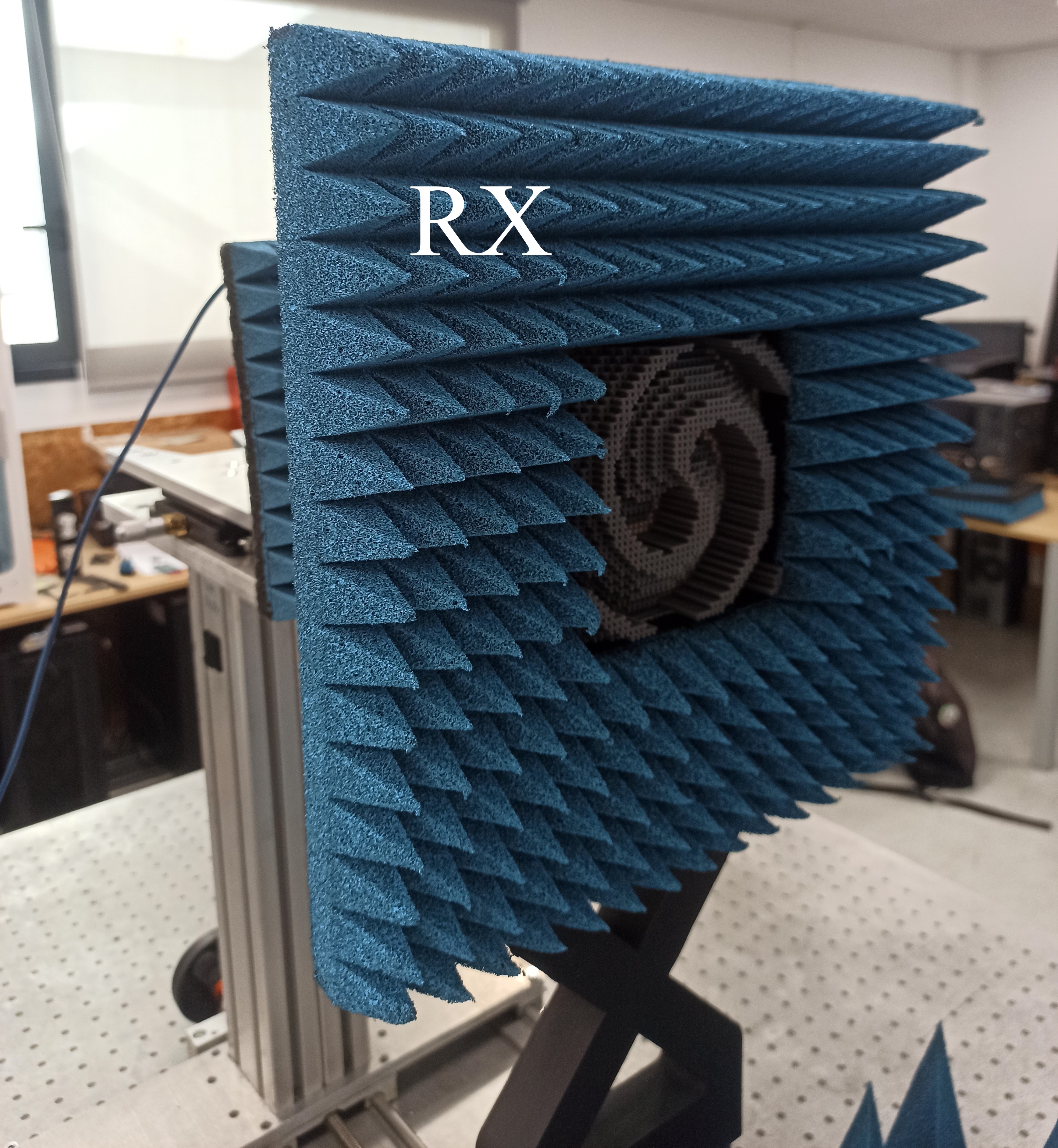}\label{fig:receiver_comms}}
\caption{Pictures of the experimental setup for OAM-based communications. (a) Panoramic. (b) TX side. (c) RX side.}
\label{Fig:OAM_comms}
 \end{figure}

Several pictures of the setup are presented in Fig. \ref{Fig:OAM_comms}, where both the TX and RX\footnote{Henceforth, TX and RX encompass both the feed and the transmitarray at the transmitter and receiver sides.} are covered with absorbents to avoid possible diffraction effects with the edges of the metasurfaces. The setup is deployed over an optical bench located in an indoor laboratory. Mechanical supports are used to ensure an accurate alignment in both horizontal and vertical dimensions. Three different TX-RX configurations are analyzed to assess the influence of orthogonality between OAMs: (i) free-space propagation, that is, with no transmitarrays; (ii) communication from $L$\:$=$\:2 TX to $L$\:$=$\:2 RX; and (iii) from $L$\:$=$\:2 Tx to $L$\:$=$\:6 Rx. Additionally, in order to evaluate the impact of defocusing on the wireless system, three different distances between TX and RX are studied: 30\:cm, 45\:cm and 60\:cm. Once the received signals were demodulated, the following performance parameters were estimated: EVM (Error Vector Magnitude), MER (Modulation Error Ratio) and BER (Bit Error Rate). The obtained results are summarized in Table \ref{tab:macro-tabla} for each configuration and TX-RX distance.

\begin{figure*}[t!]
\centering   
\subfigure[]{\includegraphics[width=0.65\columnwidth]{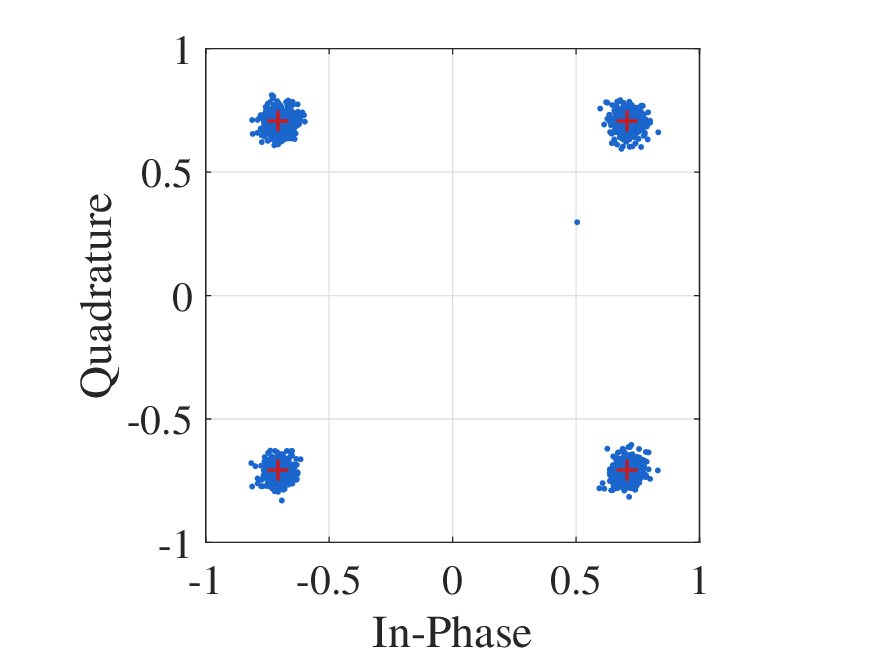}\label{fig:QPSK_free}}
\subfigure[]{\includegraphics[width=0.65\columnwidth]{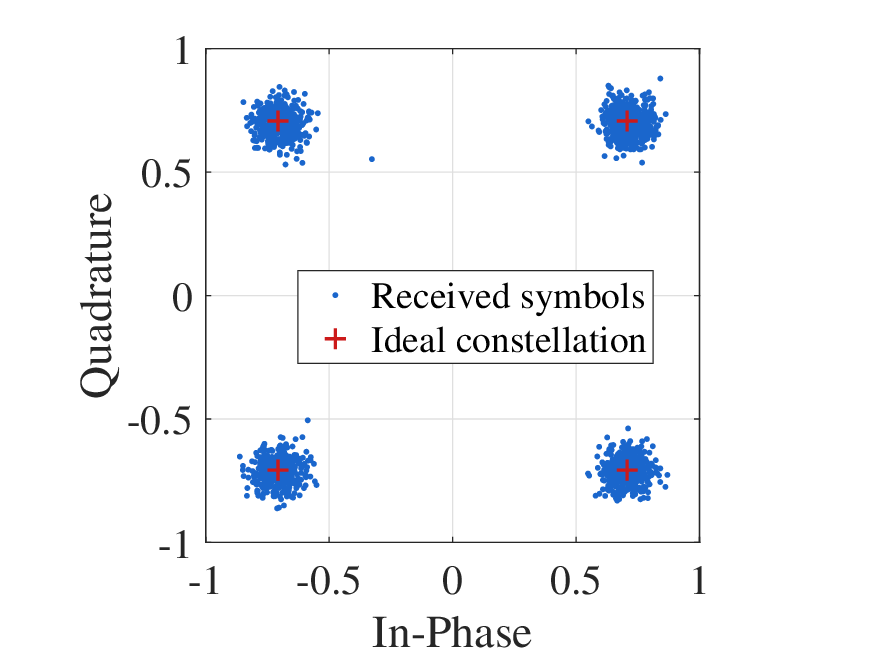}\label{fig:QPSK_l2l2}}
\subfigure[]{\includegraphics[width=0.65\columnwidth]{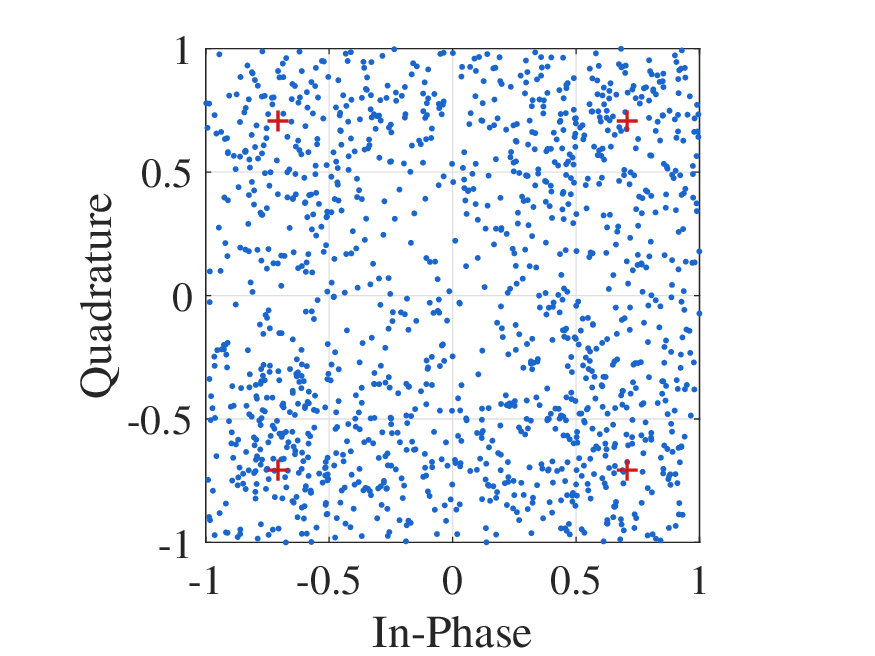}\label{fig:QPSK_l2l6}}
\caption{Constellation diagrams of the received QPSK symbols (depicted in blue) at a Tx-Rx distance of 30\:cm compared to the ideal ones (depicted in red) for the configurations: (a) free-space, (b) L2 to L2 and (c) L2 to L6.}
\label{fig:QPSK}
\end{figure*}

In the case of a distance of 30\:cm (focus distance of the $L$\:$=$\:2 OAM), similar performance is observed for the free-space and L2 to L2 configurations. In particular, EVM values of 5.1\% and 7.4\% are respectively obtained which indicate low distortion. These are aligned with the MER values of 25.9\:dB and 22.7\:dB, providing BERs lower than 10\textsuperscript{$-$6}. Conversely, for the L2 to L6 configuration (communication between orthogonal modes), the performance notably decays obtaining an EVM of 51.9\% and a MER of 5.7\:dB; that is, about 20\:dB lower than for the other configurations. These parameters lead to a BER in the order of 10\textsuperscript{$-$2}, which is unacceptable for a good quality communication. The constellation diagrams for the three configurations are presented in Fig. \ref{fig:QPSK}, where the received symbols (depicted as blue points) are compared to the ideal QPSK ones (red crosses). As aforementioned, it can be seen that free-space and L2 to L2 configurations provide a similar performance with a low dispersion respect to the ideal constellation points. In contrast, the L2 to L6 configuration presents a high dispersion in the received points which is translated into low performance parameters due to the orthogonality between modes.

On the other hand, when the TX-RX distance increases, the communication performance deteriorates for the free-space configuration. At a separation of 45\:cm, the EVM is incremented in a 15\%-rate whereas the MER gets a reduction of more than 10\:dB. Consequently, the BER rises one order of magnitude, up to to 10\textsuperscript{$-$5}. Similar performance is observed for a distance of 60\:cm. That is not the case for the L2 to L2 configuration, where performance remains stable with an EVM of 6.0\% and a MER of 24.5\:dB for a distance of 45\:cm. However, the communication notably deteriorates when TX and RX are separated 60\:cm reaching a BER in the order of 10\textsuperscript{$-$3}. These results reveals that the OAM wave is capable of transmitting information farther than the design focus distance because of the tolerance given by the convergence of the OAM. That is, since the OAM converges to a point while propagating, a certain region can be defined where the demodulation  without notable errors is still possible. Lastly, as expected, the orthogonal configuration L2 to L6 provides low performance for all TX-RX distances. Nevertheless, it seems that communication quality slightly increases with a BER in the order of 10\textsuperscript{$-$2} at 30\:cm to 10\textsuperscript{$-$3} at 60\:cm. This is a consequence of the loss of orthogonality for larger TX-RX distances where communication approaches to the free-space conditions. This condition arises from beam divergence. Due to the inherent characteristics of OAM beams and the finite aperture of the transmitting antenna, the incident wavefront reaches the receiving metasurface with a non-negligible spatial spread, resulting in partial interception of the radiated energy.

In summary, the conducted field test reveals that communication based on non-orthogonal OAMs provides similar performance than free-space propagation. Moreover, even though the OAM is designed to focus at a certain distance, from a practical view it can be properly received within a tolerance region. On the other hand, when orthogonal modes are placed in TX and RX (e.g. L2 to L6 configuration), communication is completely invalidated. The orthogonality notably increases the dispersion on the received constellation respect to the free-space conditions. These insights can be directly applied in communication techniques such as OAM multiplexation or physical-layer security.
These conclusions are specific to the proposed system configuration; however, alternative implementations without focusing capabilities can be designed to support long-distance (far-field) communication links. Moreover, by combining near-field focusing and non-focusing architectures, multiple physical-layer configurations can be realized within a single communication system.

\begin{table*}[t]
    \caption{Performance parameters of the OAM-based communication tests for the different \\ configurations and Tx-Rx distances. Results are given in the form: mean (standard deviation)}
    \centering
    \renewcommand{\arraystretch}{1.25}
    \label{tab:macro-tabla}
    \begin{tabular}{|c|c||c|c|c|}
        \cline{2-5}
        \multicolumn{1}{c|}{} & \textbf{Distance [cm]} & \textbf{EVM [\%]} & \textbf{MER [dB]} & \textbf{BER} \\ \hline
        & 30 & 5.054 (0.064) & 25.932 (0.111) & $<$ 10\textsuperscript{$-$6} \\
        \textbf{Free-space} & 45 & 20.195 (0.411) & 13.897 (0.176) & 1.666 (6.630) $\times$ 10\textsuperscript{$-$5} \\
        & 60 & 20.128 (0.365) & 13.925 (0.158) & 1.111 (6.747) $\times$ 10\textsuperscript{$-$5} \\ \hline
        & 30 & 7.350 (0.093) & 22.666 (0.110) & $<$ 10\textsuperscript{$-$6} \\
        \textbf{L2 to L2} & 45 & 5.972 (0.070) & 24.479 (0.102) & $<$ 10\textsuperscript{$-$6} \\
        & 60 & 54.114 (2.733) & 5.336 (0.203) & 5.01 (5.50) $\times$ 10\textsuperscript{$-$3} \\ \hline
        & 30 & 51.926 (3.299) & 5.706 (0.451) & 4.18 (0.71) $\times$ 10\textsuperscript{$-$2} \\
        \textbf{L2 to L6} & 45 & 47.185 (1.103) & 6.526 (0.203) & 2.76 (0.43) $\times$ 10\textsuperscript{$-$2} \\
        & 60 & 33.925 (0.744) & 9.392 (0.190) & 3.29 (1.01) $\times$ 10\textsuperscript{$-$3} \\ \hline
    \end{tabular}
\end{table*}

\section{Conclusions}

This study examines the NF indoor communications capabilities of OAM waves in the FR2 frequency band. For that purpose, a fully dielectric unit cell is proposed and characterized through a semi-analytical approach supported by a circuit model. The limitations of this unit cell are studied, with particular emphasis on its response under the oblique incidence. After several tests, where the unit cell showcase its remarkable robustness even up to an incidence angle of 30º, a series of OAM-generating TAs are subsequently designed, fabricated, and prepared for further testing. Through them, an intensive study on the establishment of communications has been carried out. On the one hand, it is shown that orthogonality between OAMs is a good coding element, in such a way that the emitted energy is well captured if and only if the receiver has the same topological charge as the transmitter. It is also evident that the presence of objects may not alter the communication thanks to the intrinsic properties of the type of wave studied. Experimental results show a $15$\ dB difference when an object is placed at the system center and the receiver operates in the same mode as the transmitter, compared to a different-mode configuration. On the other hand, the OAM-based communication field-test reveals that this TA prototype is capable to provide similar performance as free-space at even larger distances thanks to the convergence of the OAM wave, obtaining a BER lower than 10\textsuperscript{$-6$}. In contrast, when orthogonal modes are used in TX and RX, communication performance completely decays with an increase in BER to the order of 10\textsuperscript{$-$2}. This opens the door to consider the use of these systems for communications in short-range scenarios enabling innovative techniques such as multiplexation based on OAM modes or physical-layer security supported by the orthogonality.

\end{document}